\documentclass[epjST]{svjour}
\usepackage[T1]{fontenc}
\usepackage[latin9]{inputenc}
\usepackage{amsmath}
\usepackage{amssymb}
\usepackage{graphicx}
\usepackage{bbm}
\usepackage[numbers]{natbib}
\renewcommand{\vec}[1]{\mathbf{#1}}
\usepackage{braket}
\usepackage[draft]{changes}

\begin{document}
	\title{How to Win Friends and Influence Functionals: Deducing Stochasticity From Deterministic Dynamics}
	\author{Gerard McCaul \inst{1}\fnmsep\thanks{\email{gmccaul@tulane.edu}} \and Denys. I. Bondar \inst{1}\fnmsep\thanks{\email{dbondar@tulane.edu}} }

    \date{\today}

\abstract{
		 The longstanding question of how stochastic behaviour arises from deterministic Hamiltonian dynamics is of great importance, and any truly holistic theory must be capable of describing this transition. In this review, we introduce the \emph{influence functional} formalism in both the quantum and classical regimes. Using this technique, we demonstrate how irreversible behaviour arises generically from the reduced microscopic dynamics of a system-environment amalgam. The influence functional is then used to rigorously derive stochastic equations of motion from a microscopic Hamiltonian. In this method stochastic terms are not identified heuristically, but instead arise from an exact mapping only available in the path-integral formalism. The interpretability of the individual stochastic trajectories arising from the mapping is also discussed. As a consequence of these results, we are also able to show that the proper classical limit of stochastic quantum dynamics corresponds non-trivially to a generalised Langevin equation derived with the classical influence functional. This provides a further unifying link between open quantum systems and their classical equivalent, highlighting the utility of influence functionals and their potential as a tool in both fundamental and applied research.}
	\maketitle
	
	\section{Introduction\label{sec:Introduction}}
	The predictive power of physics rests on the presumption of universal
	laws. These include global spatial and temporal symmetries which demand
	momentum and energy conservation \cite{Goldstein}, while time reversal
	symmetry arises as a consequence of Hamiltonian dynamics \cite{arnold1989mathematical}.
	Problematically however, we do not see the conservation implied by
	fundamental symmetries in mundane experience. Energy leaks, structure
	deteriorates, and lifetimes (both correlative and biological) are
	finite. This is an altogether antique notion - ``all human things
	are subject to decay/And when fate summons, monarchs must obey''
	\cite{9780675092999} - and it must be accounted for in physical theories. In order to model systems displaying the characteristics of dissipation
	and fluctuation familiar to us in everyday life, one must use a \emph{statistical} description. 
	
Stochastic descriptions of physics were originally motivated by a desire to prove the existence of atoms, as Einstein's description of Brownian motion was framed as an experimentally observable consequence of an atomistic picture \cite{Cohen2005}. This result inspired a proliferation of stochastic methods in physics, with a variety of formalisms used to describe them \cite{10.1143/PTP.33.423,PhysRev.124.983}. In particular, stochastic thermodynamics \cite{Seifert2012} predicts thermodynamic behaviour at both macro and microscopic scales \cite{PhysRevX.7.011008} using microscopic stochastic models. This approach has been enormously successful, generalising the laws of thermodynamics \cite{PhysRevE.60.2721,PhysRevLett.78.2690} and providing rich links with information theory \cite{PhysRevLett.121.030605}. 

One feature of stochastic theories is their ability to capture the aforementioned phenomena of dissipation and fluctuation, which renders them intrinsically irreversible. This approach stands in marked contrast to the presumption of global spatial and temporal symmetries in the microscopic description of physical systems \cite{Goldstein}. The apparent contradiction between statistical and microscopic mechanics is made explicit by the Loschmidt paradox \cite{Timearrow}, raising the question of how irreversible behaviour may arise from reversible dynamics. This is a problem of fundamental importance, and its ultimate resolution requires a rigorous mapping from a microscopic Hamiltonian to effective irreversible dynamics. 

Here, we provide exact derivations of these mappings, for both quantum and classical systems. In the former case a powerful path integral technique known as the Feynman-Vernon influence functional \cite{Feynman-Vernon-1963} is used. This formalism allows one to characterise the effect of an environmental coupling to an open system without reference to the environment. It is a powerful and flexible formalism that can be used to attack the problem of open quantum systems, yielding a number of both exact \cite{Kleinertbook,KLEINERT1995, Tsusaka1999}  and approximate \cite{Smith1987,Makri1989,Allinger1989, Bhadra2016,McDowell2000} results. Influence functionals have been deployed in the study of both real and imaginary time path integrals. In real time, influence functionals have been used to rigorously derive quantum Langevin  equations \cite{Caldeira1983,Ford-Kac-JST-1987,Gardiner-1988,Sebastian1981,Leggett1987,van_Kampen-1997},  stochastic Schr\"{o}dinger \cite{Orth2013,Orth2010}, quantum Smoluchowski \cite{Ankerhold2001,Maier2010} and Liouville-von Neumann \cite{Stockburger2002, Stockburger2004, Stockburger2017} equations, as well as quasiadiabatic path integrals \cite{Nalbach2009}. Additionally, the models derived via influence functionals have also been used successfully in both imaginary and real time numerical simulations of dissipative systems \cite{Banerjee2015,Makri2014,Makri1998,Dattani2012,Habershon2013,Herrero2014,Wang2007, ourpaper2}.

\begin{figure}
\includegraphics[width=\columnwidth]{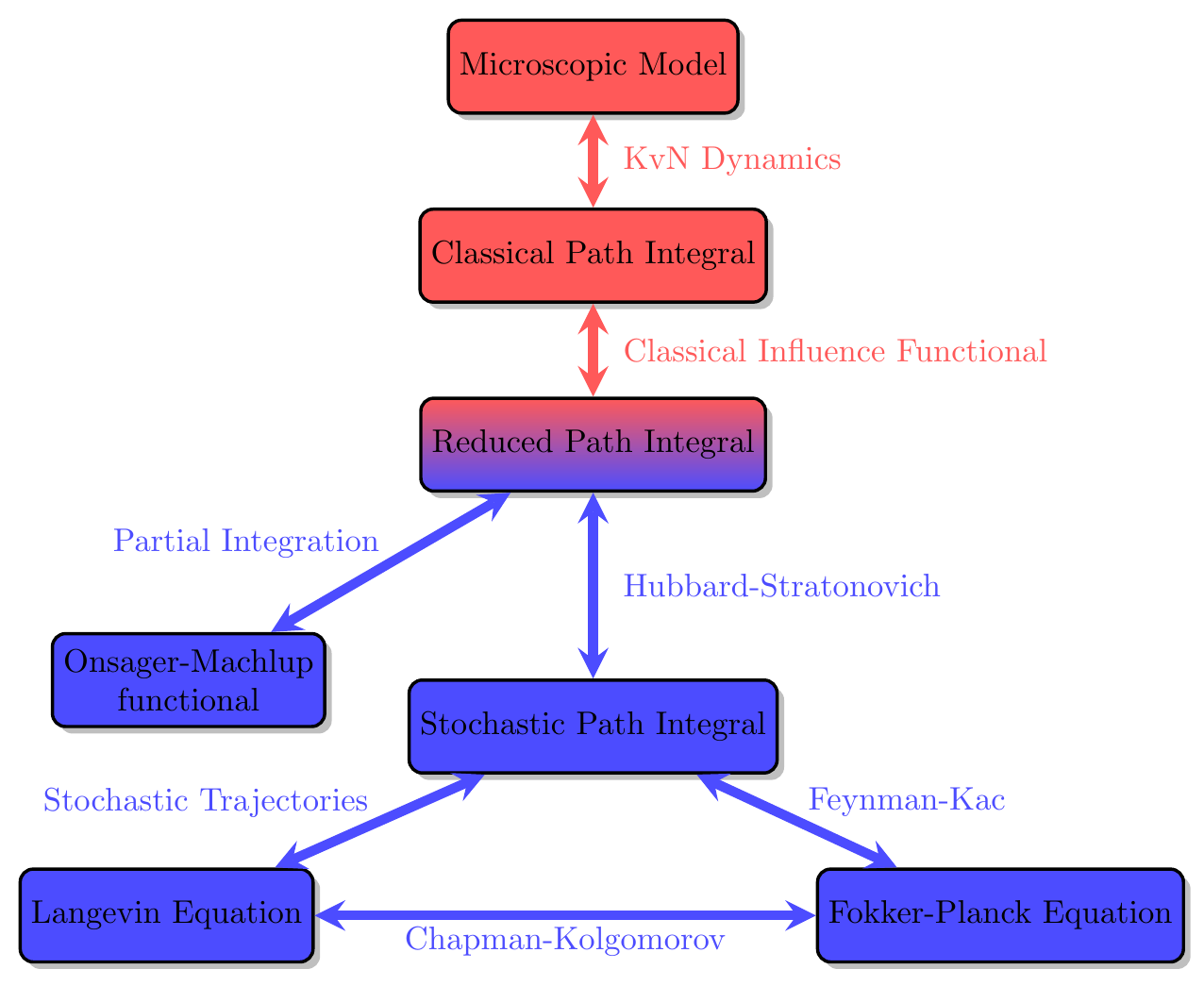} \caption{Schematic of the transition from deterministic to stochastic dynamics described in this paper. Using the Classical Influence Functional  (CIF), one may formally transition from a deterministic microscopic model to any of a number of stochastic descriptions. \label{fig:influenceshematic}}
\end{figure}

Given the tremendous utility of the influence functional in a quantum mechanical context, the development of a classical analogue is of great importance, both as a method to explore fundamental physics and as a tool for practical calculation. Alongside the well-established quantum influence functional, it is possible to derive a \emph{Classical Influence Functional} (CIF) that may be applied to Hilbert space representations of classical mechanics. This can then be mapped to other methods for modelling irreversible processes, such as the Fokker-Planck equation \citep{risken1989}, the stochastic path integral \citep{Weber_2017}, or the Langevin equation \citep{Langevintranslate}, as seen in Fig. \ref{fig:influenceshematic}. 

We now outline the structure of the paper, beginning in Sec. \ref{sec:KvN} with the essential prerequisite for the CIF, the representation of classical dynamics in Hilbert space using \emph{Koopman-von Neumann} (KvN) dynamics. This yields the most transparent comparison between quantum and classical systems, recasting classical dynamics in the same formalism as quantum mechanics, and allows the classical propagator to be expressed as a path integral. The influence functional itself is introduced in Sec. \ref{sec:influence}, first for a quantum mechanical system, followed by a derivation of the classical equivalent using the KvN path integral. Section \ref{sec:EOMs} returns to the original motivation of this review, using both forms of influence functional to derive effective, irreversible equations of motion directly from Hamiltonian dynamics. The CIF is then used to establish the proper classical limit of quantum stochastic dynamics in Sec. \ref{classicallimit}, along with a discussion of the physical interpretation of stochastic trajectories. Finally, we close the paper with a discussion of the results presented here.

	\section{Koopman-von Neumann dynamics \label{sec:KvN}}
We now introduce the KvN formalism for classical mechanics. This is in a sense the adjoint to formulations of quantum mechanics in phase space \cite{PhysRev.40.749, Baker1958,Curtright2014,Groenewold1946}. While the latter theories are ``classicalised'' descriptions of quantum phenomena, KvN mechanics casts classical physics in a quantum language, by reformulating it in a Hilbert space formalism \cite{Koopman315}. This operational formalism underlies ergodic theory \cite{ReedSimon2}, and is a natural platform to model quantum-classical hybrid systems  \cite{Sudarshan1976,Viennot_2018,Bondarquantumclassical}. KvN dynamics has also been employed to establish a classical speed limit for dynamics \cite{PhysRevLett.120.070402}, as well as enabling an alternate formulation of classical electrodynamics \cite{Hilbertelectrodynamics}. Furthermore, using KvN dynamics it is even possible to combine classical and quantum dynamics in a unified framework known as operational dynamical modeling \cite{Bondar2012,Bondar2013}. 

In terms of application, the KvN formalism has been used productively to study linear representations of non-linear
dynamics \cite{Koopman-nonlinear}, dissipative behaviour
\cite{Chruscinski2006}, and entropy conservation \cite{PhysRevE.99.062121}. It has also been applied in various industrial contexts \cite{mezic2005spectral,budivsic2012applied,Koopman2020} and to  analysis of the time dependent harmonic oscillator \cite{RamosPrieto2018}. In the current context, KvN's main utility is that it will allow for the direct importation of the quantum mechanical techniques that underlie the influence functional. 
	
	\subsection{The Koopman Operator}
	KvN is a Hilbert space theory, so to begin with let us define its
essential characteristics. In its functional form, a Hilbert space
consists of the set of functions $L^{2}$$\left(\mathcal{P},{\rm d}\mu\right)$:

\begin{equation}
L^{2}\left(\mathcal{P},{\rm d}\mu\right)=\left\{ \phi\,:\,\mathcal{P}\to\mathbb{C}\left|\int{\rm d}\mu\ \left|\phi\right|^{2}<\infty\right.\right\} 
\end{equation}
i.e. the set of all functions on a space $\mathcal{P}$ that are square
integrable with a measure ${\rm d}\mu$. The other necessary ingredient
in a Hilbert space is the definition of an inner product

\begin{equation}
\left\langle \phi\left|\psi\right.\right\rangle =\int{\rm d}\mu\ \phi^{*}\psi
\end{equation}
for $\phi,\psi\in L^{2}\left(\mathcal{P},{\rm d}\mu\right)$. Physics is introduced to this formalism by interpreting the elements
of $L^{2}$ as probability density amplitudes. In addition, observables
are associated with Hermitian operators and states obey the Born rule.
The probability distribution for a state is therefore the square of
its wavefunction $\rho=\left|\psi^{2}\right|$. In addition, Stones'
theorem guarantees there exists a one-parameter, continuous group
of unitary transformations on $L^{2}$ of the form \cite{Stonetheorem}:
\begin{equation}
\hat{U}\left(t\right)={\rm e}^{i\hat{A}t}\label{eq:TheoremStones}
\end{equation}
where $\hat{A}$ is a unique \emph{self-adjoint} operator.
This family of transformations is interpreted as time evolution and
leads to the following differential equation:
\begin{equation}
\dot{\psi}=i\hat{A}\psi
\end{equation}
So far, this is identical to quantum mechanics. The key distinction
between quantum and Koopman dynamics is the way elements of $L^{2}$
are evolved in time. Specifying the form of $\hat{A}$ adds physics
to the formalism, and requires both the imposition of the Ehrenfest
theorems, \emph{and} a fundamental commutation relation \cite{Bondar2012a}.
The choice of commutation relation is the sole distinction between
Koopman dynamics and quantum mechanics \cite{Bondar2012}. To see this, consider the  Ehrenfest theorems:
\begin{align}
\frac{{\rm d}\left\langle \hat{x}\right\rangle }{{\rm d}t}&=\left\langle \frac{\hat{p}}{m}\right\rangle, \\
\frac{{\rm d}\left\langle \hat{p}\right\rangle }{{\rm d}t}&=-\left\langle \hat{V}^{\prime}\left(x\right)\right\rangle 
\end{align}
where $\hat{V}^{\prime}=\frac{\partial V(x)}{\partial x}$ is the gradient of the system potential $V(x)$. Since these equations should hold for any state, we find the following relations for
the time generator:
\begin{align}
i\left[\hat{A},\hat{x}\right]&=\frac{\hat{p}}{m},\label{eq:firstEhrenfest} \\
i\left[\hat{A},\hat{p}\right]&=-\hat{V}^{\prime}\left(x\right)\label{eq:SecondEhrenfest}
\end{align}

In the quantum case $\left[\hat{x},\hat{p}\right]=i\hbar$. When this
is applied to Eqs. (\ref{eq:firstEhrenfest}, \ref{eq:SecondEhrenfest})
they uniquely identify the self-adjoint operator which recovers the familiar
Schr\"{o}dinger equation:

\begin{equation}
i\hbar\dot{\psi}_{{\rm qm}}=\hat{H}\psi_{{\rm qm}}
\end{equation}
In KvN mechanics $[\hat{x},\hat{p}]=0$. As a result, the
$\hat{x}$ and $\hat{p}$ operators have a common set of eigenstates.
These form an orthonormal eigenbasis, furnished with the usual relationships:
\begin{align}
	\hat{x}\left|x,p\right\rangle =& x\left|x,p\right\rangle, \ \ \left\langle x,p\left|x^{\prime},p^{\prime}\right.\right\rangle = \delta\left(x-x^{\prime}\right)\delta\left(p-p^{\prime}\right), \notag \\ 
	\hat{p}\left|x,p\right\rangle =&p\left|x,p\right\rangle, \ \ \int{\rm d}x{\rm d}p\ \left|x,p\right\rangle \left\langle x,p\right|=1.
	\end{align}
	One consequence of allowing the phase space operators to commute is
	that it is \emph{impossible} to construct an operator that satisfies Eqs. (\ref{eq:firstEhrenfest}, \ref{eq:SecondEhrenfest}) purely from $\hat{x}$ and $\hat{p}$ \cite{Mythesis}. It
	is therefore necessary to introduce two new operators, $\hat{\lambda}$
	and $\hat{\theta}$ with the commutation relations:
	\begin{align}
	\left[\hat{x},\hat{\lambda}\right]=\left[\hat{p},\hat{\theta}\right]=i, \quad
	\left[\hat{\lambda},\hat{\theta}\right]=\left[\hat{\lambda},\hat{p}\right]=\left[\hat{\theta},\hat{x}\right]=0.
	\end{align}
	
The new operators are Bopp operators \cite{AIHP_1956__15_2_81_0, doi:10.1086/288104}, and may be physically interpreted as the operational equivalent of Lagrange multipliers. Specifically, each operator acts as the Lagrange multiplier enforcing one of Hamilton's equations, an interpretation which follows from Eq. (\ref{eq:Kpathintexponent}). With these new operators, one is able to derive the propagator for classical states
	\begin{align}
	\hat{U}_{{\rm cl}}\left(t\right)={\rm e}^{-it\hat{K}},
	\end{align}
	where $\hat{K}$ is the \emph{Koopman operator}
	\begin{align}
	\hat{K}=\hat{p}\hat{\lambda}/m-\hat{V}^{\prime}\left(\hat{x}\right)\hat{\theta} .\label{eq:Koopmanoperator}
	\end{align}

	\subsection{Liouville's Theorem for KvN Classical Mechanics}
	We now show that the Koopman operator is
	consistent with more standard formulations of classical dynamics.
	Taking the evolution equation 
	\begin{align}
	i\frac{{\rm d}}{{\rm d}t}\left|\psi\right\rangle =\hat{K}\left|\psi\right\rangle 
	\end{align}
	we pick a specific representation (for more information, see appendix Sec. \ref{sec:bases}):
	\begin{align}
	\psi \equiv\psi\left(x,p\right)=&\left\langle x,p\left|\psi\right.\right\rangle, \quad \hat{x}\to x, \quad \hat{p}\to p ,\notag \\
	\hat{\lambda}\to&-i\frac{\partial }{\partial x}, \quad \hat{\theta}\to-i\frac{\partial}{\partial p},
	\end{align}
which leads to
	\begin{align}
	i\dot{\psi}=\frac{p}{m}\frac{\partial\psi}{\partial x}-V^{\prime}\left(x\right)\frac{\partial\psi}{\partial p}.\label{eq:KoopmanwavefunctionEvolution}
	\end{align}
	This evolution equation may be expressed in a more familiar form:
	\begin{align}
	\dot{\psi}= & i\hat{K}\psi=\left\{ H,\psi\right\},  \\ 
	\left\{ H,\odot\right\} = &\frac{\partial H}{\partial x}\frac{\partial\odot}{\partial p}-\frac{\partial H}{\partial p}\frac{\partial\odot}{\partial x}.
	\end{align}
	The phase space representation of the Koopman operator is the \emph{Poisson
		bracket.} The evolution equation for the classical wavefunction is
	therefore identical to that for the associated probability density
	\begin{align}
	\dot{\rho}=\left\{ H,\rho\right\} .
	\end{align}
	This fact is particularly helpful, as it means that a classical wavefunction
	and its equivalent probability density are evolved by the \emph{same}
	propagator
	\begin{align}
	U_{{\rm cl}}\left(x_{f},p_{f},t_{f};x_{i},p_{i},0\right)=\langle x_f,p_f |{\rm e}^{-it\hat{K}}|x_i,p_i\rangle, 
	\end{align}
	leading to the evolution equations:
	\begin{align}
	\psi\left(x_{f},p_{f}\right)=& \int{\rm d}x_{i}{\rm d}p_{i} U_{{\rm cl}}\left(x_{f},p_{f},t_{f};x_{i},p_{i},0\right)\psi\left(x_{i},p_{i}\right), \\
	\rho\left(x_{f},p_{f}\right) =& \left|\psi\left(x_{f},p_{f}\right)\right|^{2}\nonumber \\ 
	=& \int{\rm d}x_{i}{\rm d}p_{i} U_{{\rm cl}}\left(x_{f},p_{f},t_{f};x_{i},p_{i},0\right)\rho\left(x_{i},p_{i}\right).
	\end{align}
	
	\subsection{KvN Path Integral}
	We close this section with a discussion of path integral formulations of KvN. It is possible in this formalism to construct both deterministic and stochastic classical path integrals \cite{Gozzi2014, 1505.06391}, including generalisations with geometric forms \cite{Gozzi1994}. These path integral formulations may be usefully applied with classical many-body diagrammatic methods {\cite{Liboff}}, but in our case, they shall be used to derive the influence functional. 
	
	A full derivation of the KvN path integral is available in Appendix \ref{pathintegral}, and we quote the result here. The classical propagator (dropping its arguments for brevity) may be expressed as
		\begin{align}
		U_{{\rm cl}} =&  
		\int \mathcal{D}x(t) \mathcal{D}\theta (t) \  {\rm e}^{i\int_{0}^{t_{f}}{\rm d}t \ \theta\left(t\right)\left[m\ddot{x}\left(t\right)+V^{\prime}\left(x(t)\right)\right]} ,\label{eq:Koopmanpropagator}
		\end{align}
with a functional measure given by 
		\begin{align}
		\mathcal{D}x (t) \mathcal{D}\theta (t) =\lim_{N\to\infty}\left(\frac{m}{2\pi\Delta}\right)^{N}\prod_{n=1}^{N}{\rm d}x_{n}{\rm d}\theta_{n} \label{eq:kvnmeasure}
		\end{align}
where $\Delta$ is the time-step resulting from the discretisation of the propagator. It is easy to see that the exponent in the classical path integral is a delta functional which enforces precisely the classical equations of motion, where the kernel of the exponent is the KvN equivalent
		to the action in the quantum path integral. If we consider the limit of localised probability distributions $\rho_{0}\left(x_{i},p_{i}\right)=\delta\left(x_{i}-x_{0}\right)\delta\left(p_{i}-p_{0}\right)$, the distribution at later times is described by
		\begin{align}
		\rho\left(x_{f},p_{f},t_{f}\right)=&\int{\rm d}x_{i}{\rm d}p_{i} \ U_{{\rm cl}} \rho_{0}\left(x_{i},p_{i}\right) \nonumber \\
		=&\int{\rm d}x_{i}{\rm d}p_{i}\mathcal{D}x(t)\ \rho_{0}\left(x_{i},p_{i}\right)\delta\left[m\ddot{x}\left(t\right)+V^{\prime}\left(x,t\right)\right] \nonumber \\ =&\delta\big(x_{f}-x_{{\rm cl}}\left(t_{f}\right)\big)\delta\big(\dot{x}_{f}-\dot{x}_{{\rm cl}}\left(t_{f}\right)\big).
		\end{align}
Hence, the particle remains localised with its trajectory $x_{{\rm cl}}\left(t\right)$
	described by the classical equation of motion. The KvN propagator
	in this special case is simply a formally excessive
	representation of single-particle Newtonian mechanics.
	Clearly, applying this formalism to single-particle classical mechanics recovers well known results, but by expressing the composite of
	an open system and its environment in this form, we are able to construct
	an influence functional to integrate out the environment explicitly.

	\section{Influence Functionals \label{sec:influence}}
\begin{figure}
\centering
\includegraphics[height=4cm]{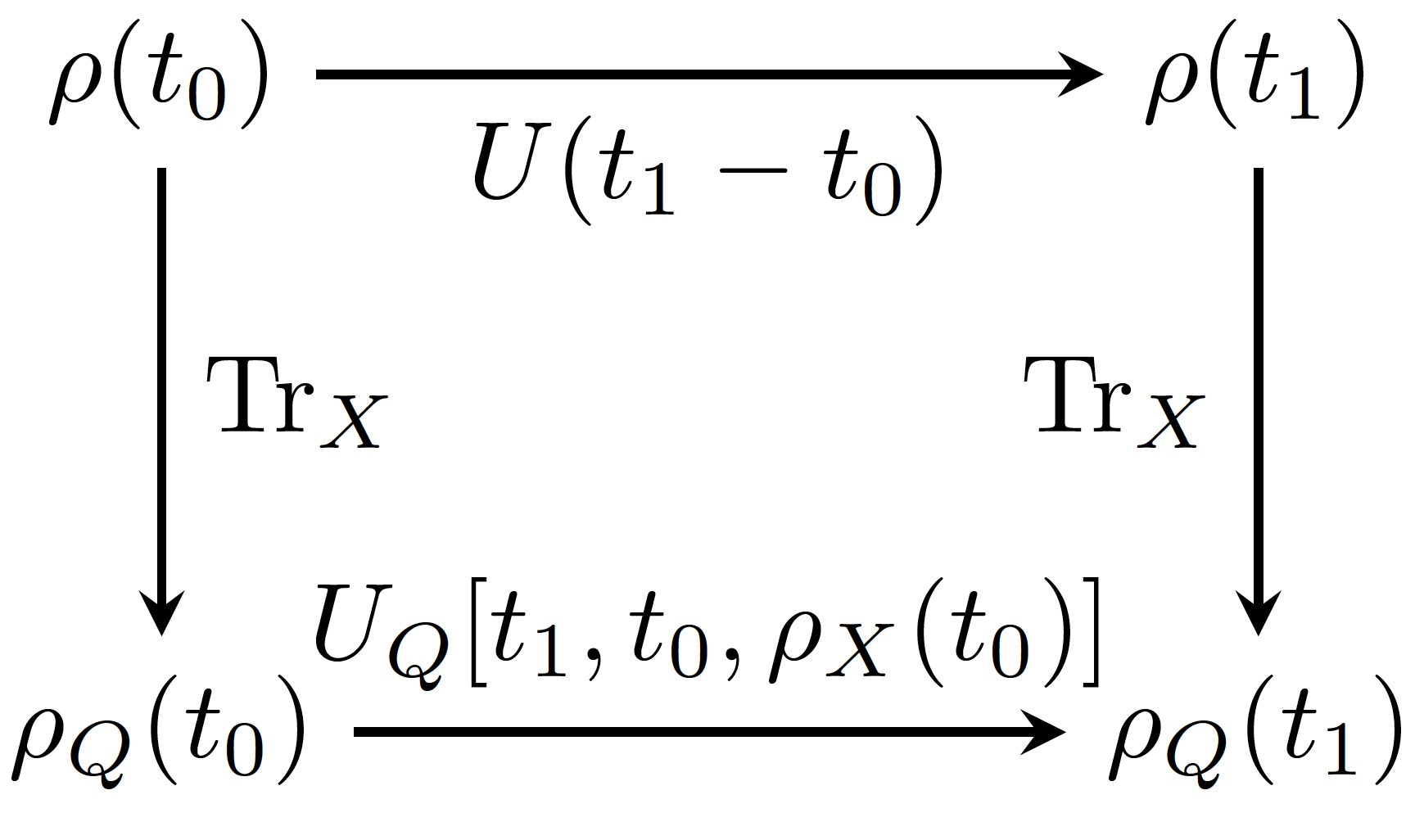} \caption{In both the classical and quantum evolutions, the composite system + environment $\rho$ (where $\rho$ may represent either a density matrix or probability density, depending on the framework) may be unitarily evolved, before tracing over the environment. Alternately, one may use influence functionals to describe an effective propagator explicitly dependent on the initial environment $\rho_X (t_0)$ which directly evolves the reduced system $\rho_Q$.\label{fig:reducedevoschematic}}
\end{figure}

	In this section we detail the construction of influence functionals, which allow one to re-express a many particle problem in terms of a modified one-body equation. For both the quantum and classical cases, the aim is to produce an \emph{effective propagator} $U_Q$ that describes the evolution of a reduced system. The role of this propagator is shown in Fig. \ref{fig:reducedevoschematic}, describing the evolution of an open system without reference to the environment.  We begin with the quantum case, the Feynman-Vernon influence functional \cite{Feynman-Vernon-1963}.
	\subsection{Quantum Influence Functional}

	Consider an open system $Q$ and an environment $X$ respectively characterised by
	collective coordinates $q$ and $x$, with an interaction $\hat{H}_I$.
	The total Hamiltonian is described by \cite{Petruccione-open-systems-book}
	\begin{align}
	\hat{H}_{{\rm tot}}=\hat{H}_{Q}+\hat{H}_{X}+\hat{H}_{I}.\label{eq:htotschematic}
	\end{align}
The fundamental object of interest for open quantum systems is the
\emph{density matrix}, defined as 
\begin{equation}
\hat{\rho}=\sum_{\alpha}p_{\alpha}\left|\psi_{\alpha}\left\rangle \right\langle \psi_{\alpha}\right|.
\end{equation}
The density matrix is a statistical mixture of pure states $\psi_{\alpha}$,
and generalises the notion of a quantum state to systems also governed
by classical probability distributions. Typically, this is necessitated
by the need to describe a thermal system, where the energy eigenstates
are weighted by the Gibbs distribution. Thus, a common density matrix
is the canonical density matrix:
\begin{equation}
\hat{\rho}_{\beta}=\frac{1}{Z_{\beta}}\textrm{e}^{-\beta\hat{H}}=\frac{1}{Z_{\beta}}\sum_{\alpha}{\rm e}^{-\beta E_{\alpha}}\left|E_{\alpha}\left\rangle \right\langle E_{\alpha}\right|
\end{equation}
\begin{equation}
Z_{\beta}=\sum_{\alpha}\textrm{e}^{-\beta E_{\alpha}}.
\end{equation}
 The relationship of a density matrix to an expectation is easily
verified,
\begin{equation}
\left\langle \hat{A}\right\rangle ={\rm Tr}\left[\hat{\rho}\hat{A}\right]=\int\mathrm{d}q\ A\left(q\right)\rho\left(q,q\right)
\end{equation}
where the final equality expresses the density matrix in a specific
basis:
\begin{equation}
\rho\left(q,q^{\prime}\right)=\sum_{\alpha}p_{\alpha}\big<q'\big|\psi_{\alpha}\big>\big<\psi_{\alpha}\big|q\big>.
\end{equation}

Finally, as we will be concerned with the evolution of open systems in both the classical and quantum case, it is worth taking a moment to consider the relationship between the probability distribution $\rho(q,p)$ and the classical limit of the quantum density matrix $\hat{\rho}$. To do so, it is most instructive to consider the description of the  density matrix in phase space using the Wigner quasi-probability distribution
\citep{Case2008} $W(q,p)$:
\begin{equation}
W\left(q,p\right)=\frac{1}{2\pi\hbar}\int{\rm d}y\ {\rm e}^{-\frac{i}{\hbar}p y}\rho\left(q+\frac{y}{2},q-\frac{y}{2}\right)
\end{equation}
This object is entirely equivalent to the density matrix \cite{Baker1958,Curtright2014,Groenewold1946}, but its representation in phase space allows for a more transparent classical limit $\hbar \to 0$. While there are some formidable subtleties to this limit \citep{Case2008}, one finds that for mixed states the probability density is recovered \cite{PhysRevE.99.062121}, while for pure states we obtain the classical wavefunction \citep{Bondar2013a}. 

	Let us now say we are only interested in the dynamics of the open system $Q$. The expectation of an operator $\hat{A}$
	acting only on the $Q$ subsystem is
	\begin{align}
	\left\langle \hat{A}\right\rangle =\int\mathrm{d}q\mathrm{d}q'\mathrm{d}x\mathrm{d}x'\ \rho(q,x;q',x';t)A(q,q')\delta\left(x-x^{\prime}\right).\label{eq:reducedexpectation}
	\end{align}
	This expression can be simplified by defining a \emph{reduced }density
	matrix $\rho_{Q}(q;q';t)$ which describes subsystem
	$Q$ by tracing out the environment $X$:
	\begin{align}
	\rho_{Q}(q;q';t)&=\int\rho(q,x;q',x';t)\delta\left(x-x^{\prime}\right)\,\mathrm{d}x{\rm d}x^{\prime}.\label{eq: reduceddensity} \\
	\implies \left\langle \hat{A}\right\rangle &=\int\mathrm{d}q\mathrm{d}q' \rho_Q(q;q';t)A(q,q')={\rm Tr}\left[\hat{\rho}_Q(t)\hat{A}\right].
	\end{align}
	Additionally, when we incorporate time evolution, the density matrix
	at time $t_{f}$ is
	\begin{align}
	\rho(q,x;q',x';t_{f})&= \int  \mathrm{d}q_{0}\mathrm{d}q'_{0}\mathrm{d}x_{0}\mathrm{d}x'_{0} \ U(q,x;q_{0},x_{0}';t_{f})\bigg. \nonumber \\ &\bigg.\times \rho_{0}(q_{0},q'_{0};x_{0},x'_{0})U^{\dag}(q',x';q'_{0}x'_{0};t_{f})
	\end{align}
	where $\rho_{0}(q_{0},q'_{0};x_{0},x'_{0})$ is the initial density matrix at $t=0$. Notice that for density matrices there are two propagators acting
	on the unprimed and primed coordinates at either side of the density
	matrix, which can be interpreted as forward and reversed time trajectories
	respectively \cite{DemichevChaiChian}. If we now insert the quantum path integral representation for the propagators
	we obtain \cite{TechniquesApplicationsPathIntegration}:
	\begin{equation}
	\rho(q,x;q',x';t_{f})=\int\mathcal{D}Q\mathcal{D}X\ {\rm e}^{i\frac{S_{{\rm tot}}}{\hbar}}\rho(q_{0},q'_{0};x_{0},x'_{0};t_{f}),
	\end{equation}
	where in the interests of concision we have made the abbreviations:
	\begin{align}
	\mathcal{D}Q=&\mathrm{d}q_{0}\mathrm{d}q'_{0}\,\mathcal{\mathcal{D}}q(t)\,\mathcal{D}q'(t) \\
	\mathcal{D}X =&\mathrm{d}x_{0}\mathrm{d}x'_{0}\,\mathcal{D}x(t)\,\mathcal{D}x'(t)\\
	S_{{\rm tot}} =&S_{Q}\left[q\left(t\right)\right]-S_{Q}\left[q'\left(t\right)\right]+S_{X}\left[x\left(t\right)\right] \nonumber -S_{X}\left[x'\left(t\right)\right] \\
	&+S_{I}\left[q\left(t\right),x\left(t\right)\right]-S_{I}\left[q'\left(t\right),x'\left(t\right)\right].
	\end{align}
	$S_{Q,X}$ are the actions derived from the isolated $Q$ and $X$
	subsystem Hamiltonians, while $S_{I}$ is the component due to the
	coupling $H_{I}$. This last equality is somewhat misleading, given
	the action is a functional of both the coordinates and their time
	derivatives. The functional arguments should therefore be thought
	of purely as labels denoting whether a particular component of the
	action is due to the forward or backward propagator trajectories.
	
	Usually when calculating dynamical properties of the reduced system,
	it is assumed that the density matrix is initially in a \textit{product
		state}, that is:
	\begin{align}
	\rho_{0}(q_{0},q'_{0};x_{0},x'_{0})=\rho_{Q}(q_{0};q_{0}')\rho_{X}(x_{0};x_{0}').\label{eq:partitionassumption}
	\end{align}
	Note that it is possible to start from more general initial conditions by representing the initial state as an additional path integral in imaginary time \cite{Grabert1988,Moix2012,ourpaper}. This is an important consideration when one has strong coupling to the environment and a partitioned initial condition is inappropriate \citep{Ankerhold2001}. For the purpose of illustrating the quantum influence functional however, we will forgo this complication. 
	
	Taking an initial product state, the reduced density matrix may be represented as
		\begin{align}
		\rho_{Q}(q;q';t_f)=&\int   \mathcal{D}Q \  \mathcal{F} \left[q\left(t\right),q'\left(t\right)\right] \rho_{Q}(q_{0};q'_{0}) {\rm e}^{\frac{i}{\hbar}\left(S_{Q}\left[q\left(t\right)\right]-S_{Q}\left[q'\left(t\right)\right]\right)}.  \label{eq:reducedevolution} 
		\end{align}
		Here $\mathcal{F}\left[q\left(t\right),q'\left(t\right)\right]$ is the \emph{influence functional},
		\begin{align}
		&\mathcal{F}\left[q\left(t\right),q'\left(t\right)\right]= \int \mathcal{D}X\rho_{X}(x_{0};x'_{0})
		\bigg.{\rm e}^{\frac{i}{\hbar}S_{\mathcal{F}}} \\
		&S_{\mathcal{F}}= S_{X}\left[x\left(t\right)\right] +S_{I}\left[q\left(t\right),x\left(t\right)\right] -S_{X}\left[x'\left(t\right)\right] -S_{I}\left[q'\left(t\right),x'\left(t\right)\right],
		\end{align}
	which explicitly integrates out the $X$ system, leaving it a pure
		function of the $Q$ system coordinates. If the influence functional
		is expressed as a complex phase
		\begin{align}
		\mathcal{F}\left[q\left(t\right),q'\left(t\right)\right]={\rm e}^{\frac{i}{\hbar}\Phi\left[q\left(t\right),q'\left(t\right)\right]}, \label{eq:quantuminfluencephase}
		\end{align}
		then it is possible to describe the evolution of the $Q$ system with an \emph{effective propagator} $U_Q$:
		\begin{align}
		U_{{\rm Q}}= {\rm e}^{\frac{i}{\hbar} \left(S_{Q}\left[q\left(t\right)\right]-S_{Q}\left[q'\left(t\right)\right]+\Phi\left[q\left(t\right),q'\left(t\right)\right]\right)}. \label{eq:quantumeffpropagator}
		\end{align}
		such that the reduced density matrix evolves according to
		\begin{align}
		&\rho_{Q}(q;q';t_f)=\int\mathcal{D}Q \ U_{{\rm Q}}\rho_{Q}(q_{0};q'_{0}),
		\end{align}

If one is able to disentangle $U_{Q}$ into a product
	of the form
	\begin{align}
	U_{{\rm Q}}\left[q\left(t\right),q'\left(t'\right)\right]=\tilde{U}_{Q}\left[q\left(t\right)\right]\tilde{U}_{{\rm Q}}^{\dagger}\left[q'\left(t'\right)\right], \label{eq:quantumeffpropagatordisentangle}
	\end{align}
	then effective Hamiltonians for forward and backward evolutions can also be derived, and hence a Liouville-von Neumann like equation of motion. Such an equation captures exactly
	the dynamics of the $Q$ system, but without any reference to the
	$X$ system it is interacting with. 
	
	This is the power of the influence
	functional, as it allows for the mapping of an interacting subsystem
	to an isolated system with a modified Hamiltonian. In the context
	of open systems, the dimensionality of the environment is enormously
	large as compared to the system of interest. Being able to use the
	influence functional to characterise without approximation the effect of an environment on an
	open system is highly desirable, even if only for numerical efficiency.
	
	\subsection{Classical Influence Functional}
	
	Using the path integral KvN formulation, it is possible to directly
	import many of the results derived for the quantum path integral.
	Principal among these is the ability to describe the reduced dynamics
	of an open system + environment amalgam with an equivalent influence
	functional formalism. For a global system described with canonical
	coordinates $q,p$ and $x,k$, the total Hamiltonian may characterised as in Eq. (\ref{eq:htotschematic}), using
	\begin{align}
	H_{{Q}}=\frac{p^{2}}{2m}+V_{Q}, \quad
	H_{X}= \frac{k^{2}}{2m_{k}}+V_{X}, \quad
	H_I =& V_{QX}.
	\end{align}
	This system is initially described by the probability density 
	\begin{align}
	\rho_{0}^{{\rm tot}}=\rho_{Q}(q_{0},p_{0})\rho_{X}\left(x_{0},k_{0}, q_0, p_0\right).
	\end{align}
 where to retain full generality, the initial environment state may also depend on the open system coordinates.
 
 Using Eq. (\ref{eq:Koopmanpropagator}), the classical
	reduced probability density may be expressed in a similar manner to Eq. (\ref{eq:reducedevolution}):
	\begin{align}
	\rho_{Q}\left(q_f,p_f\right) =\int  \mathcal{D}q(t)\mathcal{D}\theta_{Q}\left(t\right)\mathrm{d}q_{0}\mathrm{d}p_{0}\ \mathcal{F}_{\rm cl} \rho_{Q}(q_{0},p_{0})  {\rm e}^{i\int_{0}^{t_{f}}{\rm d}t\ \theta_{Q}\left(t\right)\left(m\ddot{q}\left(t\right)+\frac{\partial V_{Q}}{\partial q}\right)}, \label{eq:reduceddensitywithinfluence}
	 \end{align}
where $\mathcal{F}_{\rm cl}\equiv \mathcal{F}_{\rm cl}\left[q\left(t\right),p\left(t\right),\theta_{Q}\left(t\right)\right]$ is the \emph{Classical Influence Functional} (CIF) given by
	 \begin{align}
		 \mathcal{F}_{\rm cl} =&\int  \mathrm{d}x_{0}\mathrm{d}k_{0}\mathrm{d}x_{f}{\rm d}k_{f}\mathcal{D}x(t)\mathcal{D}\theta_{X}\left(t\right)\ \rho_{X}(x_{0},k_{0},q_0, p_0) {\rm e}^{i \int_{0}^{t_{f}}{\rm d}t \ \Gamma}, \label{eq:kvninfluence}\\ 
		\Gamma =& \theta_{X}\left(t\right)\left(m_{k}\ddot{x}\left(t\right)+\frac{\partial V_{QX}}{\partial x}+\frac{\partial V_{X}}{\partial x}\right)  +\theta_{Q}\left(t\right)\frac{\partial V_{QX}}{\partial q}.
		\end{align}

Much like the quantum case, under certain circumstances an effective equation of motion may be defined from the influence functional. This is the case when it is possible to express $\mathcal{F}_{\rm cl}$ as 
	\begin{equation}
	 \mathcal{F}_{\rm cl}= {\rm e}^{i\int_{0}^{t_{f}}{\rm d}t\ \theta_{Q}\left(t\right) \chi\left[q(t),p(t)\right]} \label{influenceforeqnmotion},
	\end{equation} 
	where $\chi\left[q(t),p(t)\right]$ is an arbitrary functional of the phase space coordinates only. In this case, substituting the influence functional into Eq.\eqref{eq:reduceddensitywithinfluence}, one finds the path integral is a delta functional over paths satisfying the equation of motion
	\begin{align}
	m\ddot{q} = -\frac{\partial V_Q}{\partial q} -\chi\left[q(t),p(t)\right].
	\end{align}  
	
	\subsection{Influence Functionals and Reversibility}
Let us now consider the physical implications contained in both the quantum and classical influence functionals. First, the tracing out of the environment corresponds physically to ignorance of the environment \emph{after} the initial time $t_0$. We denote the effective propagator evolving the reduced system from $t_0$ to $t_1$  as $\hat{U}_{Q}\left[t_0,t_1, \rho_X(t_0) \right]$, where the final argument indicates the dependence on the environment state at time $t_0$, as is clear from Eq. (\ref{eq:kvninfluence}). Given that in general $\rho_{X}\left(t_0\right)\neq \rho_{X}\left(t_1\right)$, the time reversal symmetry between forwards and backwards propagations from time $t_0$ is broken, as
	\begin{equation}
	\hat{U}^{-1}_{Q}\left[t_0,t_1, \rho_X(t_0) \right] \neq \hat{U}^\dagger_{Q}\left[t_0,t_1, \rho_X(t_0) \right],
	\end{equation}  
and the inverse of the effective propagator instead depends on the environment state at the later time
\begin{equation}
\hat{U}^{-1}_{Q}\left[t_0,t_1, \rho_X(t_0) \right] = \hat{U}^\dagger_{Q}\left[t_0,t_1, \rho_X(t_1) \right].
\end{equation}
	
\begin{figure}
\centering
\includegraphics[height=6cm]{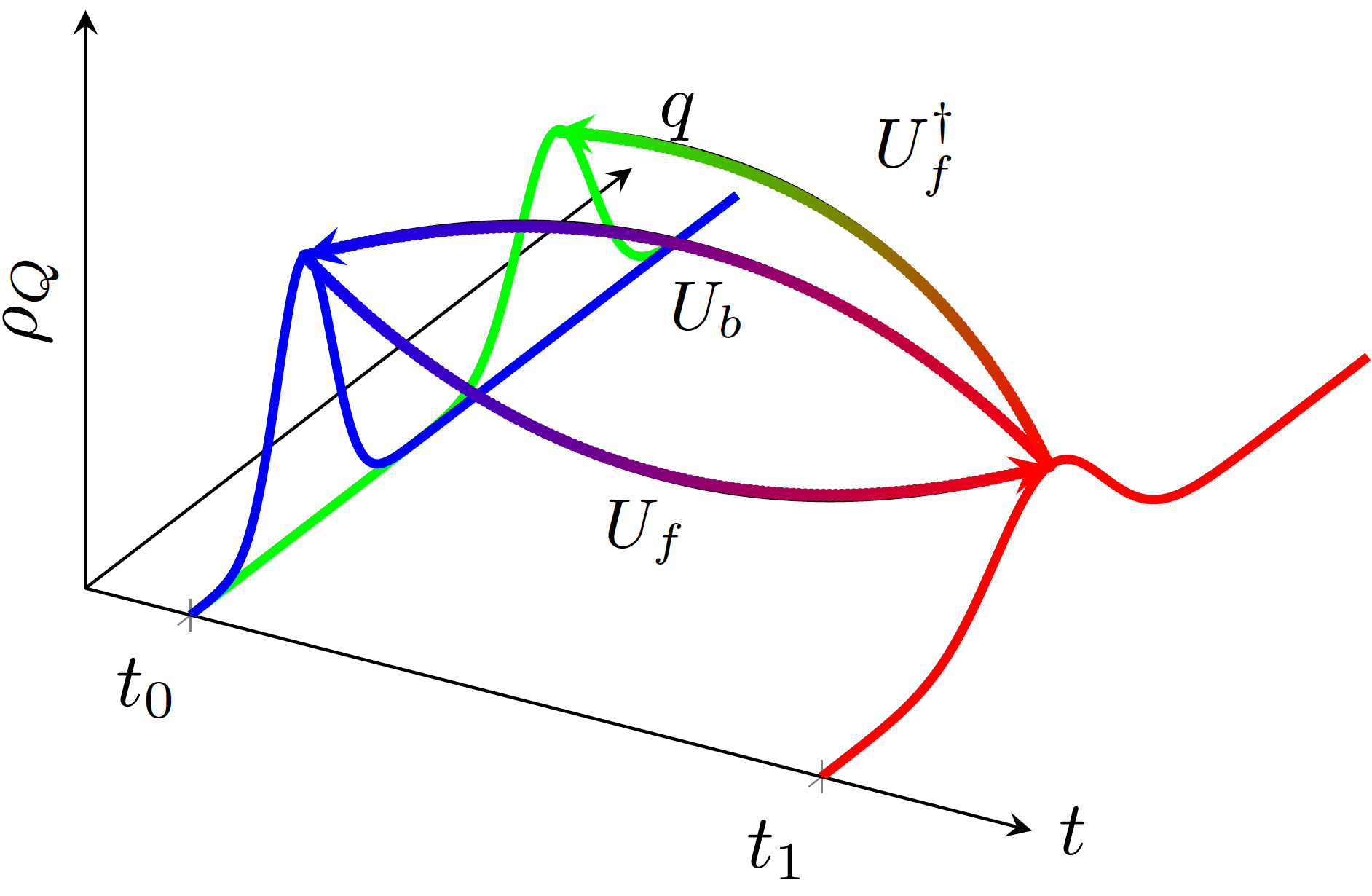} \caption{After integrating out the environment from a system, the forward propagation (from the blue initial state to the red final state) $U_f=U_Q\left[t_1, t_0, \rho_X(t_0)\right]$ is no longer a unitary transformation. The proper reversal of this propagation is $U_b= U^{-1}_f= U_Q\left[t_0, t_1, \rho_X(t_1)\right]$, while the naive assumption that $\rho_X(t_0)=\rho_X(t_1)$ corresponds to the (red to green) propagation $U^\dagger_f=U_Q\left[t_0, t_1, \rho_X(t_0)\right]$. \label{fig:forwardbackpropagations}}
\end{figure}
Thus, the evolution of the $Q$ system is \emph{itself} now dependent on initial conditions. The reverse evolution for the effective propagator from $t_1$ to $t_0$ depends on the (unknown) environment state at time $t_1$. This phenomenon is illustrated in Fig. \ref{fig:forwardbackpropagations}, where the reversing the reduced system dynamics require the state of the environment at a later time. This marginalisation of the environment at time $t_0$ represents the moment in which information about the environment is lost. Clearly, if knowledge of an unobserved part of the system (the environment) is lost at an earlier time, it is impossible to reconstruct the correct reversed effective propagator at a later time \emph{a priori}, and the observed dynamics will appear to break time reversibility. This is in effect an arrow of time, forced upon the observed system as a consequence of ignorance of the later environment state. Reversibility is only restored under the two trivial cases that the system and environment are uncoupled, or the total system begins in thermal equilibrium and the total Hamiltonian is time independent, such that $\rho_{X}\left(t_0\right) = \rho_{X}\left(t_1\right)$.  

	\section{Effective Equations of Motion \label{sec:EOMs}}
	In this section, we demonstrate how stochastic equations of motion to describe open systems in both the classical and quantum regime emerge naturally from the influence functional, and hence are derivable in a manner consistent with Hamiltonian dynamics. To do so, we consider the paradigmatic model for a system+environment amalgam, the Caldeira-Leggett (CL) Hamiltonian \citep{Caldeira1983, caldeira2014an}:
		\begin{equation}
	H_{{\rm tot}}=H_{Q}(q)+\frac{1}{2}\sum_{n}\left(k_{n}^{2}+\omega_{n}^{2}x_{n}^{2}\right)-q\sum_{n}c_{n}x_{n}.
\label{eq:CLHAM}
	\end{equation}
	This model couples an arbitrary open system with Hamiltonian $H_q$ (described by the coordinate $q)$
	to an environment of independent harmonic oscillators (with unit masses, momenta $k_{n}$,
	frequencies $\omega_{n}$, and displacement coordinates $x_{n}$),
	with each oscillator being coupled to the open system with a strength
	$c_{n}$. Very often, a counter-term  $\frac{q^{2}}{2}\sum_{n}\frac{c_{n}^{2}}{\omega_{n}^{2}}$ is added to Eq.\eqref{eq:CLHAM} to enforce translational
	invariance on the system and eliminate quasi-static effects \citep{PhysRevA.61.022107counterterm}. We have neglected this term, and any other term solely dependent on $q$, as the only effect due to these are modifications of the $Q$ system potential and distribution, which are arbitrary to begin with.
	
	The great advantage of this model is that the form of the environment and its interaction is quadratic, and the integrals one must evaluate in the influence functional are therefore Gaussian. This means that in both the quantum and classical cases, it is possible to evaluate the influence functional analytically, and derive an equation of motion. We begin with the classical case, and show that from this Hamiltonian one is able to derive a generalised Langevin equation with the CIF.
	\subsection{Classical Case: Generalised Langevin Equation \label{sec:Langevinderivation}}
	In order to derive an equation of motion, it is necessary to specify an initial condition for the extended system, which we choose to be
	\begin{align}
	\rho_{0}^{{\rm tot}}=&\rho_{Q}\left(q_{0},p_{0}\right)\rho_{\beta}\left(\vec{x}_{0},\vec{k}_{0}\right), \\ \rho_{\beta}\left(\vec{x}_{0},\vec{k}_{0}\right)=&\prod_{n}\frac{\beta\omega_{n}}{2\pi}\exp\left(-\dfrac{\beta}{2}\left(k_{0n}^{2}+\omega_{n}^{2}x_{0n}^{2}\right)\right). \label{eq:rhoebtaclassical}
	\end{align}
	This initial environment state is the Gibbs distribution for a bath of harmonic oscillators. It is actually possible to take the initial condition $\rho_{0}^{{\rm tot}}={\rm e}^{-\beta H_{{\rm tot}}}$ and include the interaction $-q\sum_{n}c_{n}x_{n}$ in $\rho_{\beta}$. In this case we would complete the square in the $\rho_{\beta}$ exponent, redefining $x_{0n}\to x_{0n}-\frac{q_{0}c_{n}}{\omega_{n}^{2}}$. This would result in an extra constant term $\beta\frac{q_{0}^{2}}{2}\sum_{n}\frac{c_{n}^{2}}{\omega_{n}^{2}}$ which could itself be cancelled by the inclusion of the counter-term mentioned above.  Critically, including interaction in the initial condition, even when it is arbitrarily strong, does not influence the structure of the derived equation of motion. 
	
	With this setup, we are able to insert the CL Hamiltonian terms into Eq. (\ref{eq:kvninfluence}). Suppressing the functional arguments of the influence functional, we obtain
		\begin{align}
		\mathcal{F}_{\rm cl} = \prod_{n} &\int \mathrm{d}x_{0n}\mathrm{d}k_{0n}\mathrm{d}x_{nf}{\rm d}k_{nf}\mathcal{D}x_{n}\left(t\right)  {\rm e}^{i\int_{0}^{t_{f}}{\rm d}t\ \theta_{Q}\left(t\right)c_{n}x_{n}\left(t\right)} \nonumber \bigg.\\ \bigg. &\times \delta\left(\ddot{x}_{n}\left(t\right)+\omega_{n}^{2}x_{n}\left(t\right)-c_{n}q\left(t\right)\right) \rho_{\beta}\left(x_{0n},k_{0n}\right),
		\end{align}
where we have replaced the integrations over $\theta_{X}$ with their equivalent delta functionals. This delta functional will force the trajectory to obey $x_{n}\left(t\right)=x_{n}^{{\rm cl}}\left(t\right)$, which solves the equation of motion $\ddot{x}_{n}^{{\rm cl}}\left(t\right)=-\omega_{n}^{2}x_{n}^{{\rm cl}}\left(t\right)+c_{n}q\left(t\right)$. Appendix \ref{sec:Oscillator} details one method of obtaining the solution quoted below
		\begin{align}
		x_{n}^{{\rm cl}}(t)=&\frac{k_{0n}}{\omega_{n}}\sin\left(\omega_{n}t\right)+x_{0n}\cos\left(\omega_{n}t\right) +\frac{c_{n}}{\omega_{n}}\int_{0}^{t}\textrm{d}t^{\prime}\,q(t^{\prime})\sin\left(\omega_{n}(t-t^{\prime})\right).
		\end{align}
		Inserting this into the influence functional
		\begin{align}
		&\mathcal{F}_{\rm cl}=\prod_{n} \int\mathrm{d}x_{0n}\mathrm{d}k_{0n} \rho_{\beta}\left(x_{0n},k_{0n}\right) \notag \\  &\times \exp\left(i\int_{0}^{t_{f}}{\rm d}t\ \theta_{Q}\left(t\right)\frac{c_{n}^{2}}{\omega_{n}}\int_{0}^{t}\textrm{d}t^{\prime}\,q(t^{\prime})\sin\left(\omega_{n}(t-t^{\prime})\right)\right)\bigg. \nonumber \\ 
		\bigg. & \times \exp\left(i\int_{0}^{t_{f}}{\rm d}t \theta_{Q}\left(t\right)c_{n}\left(\frac{k_{0n}}{\omega_{n}}\sin\left(\omega_{n}t\right)+x_{0n}\cos\left(\omega_{n}t\right)\right)\right),
		\end{align}
		and using Eq. (\ref{eq:rhoebtaclassical}) to substitute for $\rho_{\beta}$, we find that the integrals over initial positions and momenta are of a Gaussian form. Integrations over the initial phase space coordinates yields
		\begin{align}
		\int\mathrm{d}x_{0n}\ {\rm e}^{-\frac{\beta}{2}\omega_{n}^{2}\left(x_{0n}^{2}+2Ax_{0n}\right)}=&\sqrt{\frac{2\pi}{\beta}}\omega_{n}{\rm e}^{-\frac{\beta}{2}\omega_{n}^{2}A^{2}}, \\
		\int\mathrm{d}k_{0n}\   {\rm e}^{-\frac{\beta}{2}\left(k_{0n}^{2}+2Bk_{0n}\right)}=&\sqrt{\frac{2\pi}{\beta}}{\rm e}^{-\frac{\beta}{2}B^{2}},
		\end{align}
using
		\begin{align}
		A=&\frac{ic_{n}}{\beta\omega_{n}^{2}}\int_{0}^{t_{f}}{\rm d}t\ \theta_{Q}\left(t\right)\cos\left(\omega_{n}t\right), \\ B=&\frac{ic_{n}}{\beta\omega_{n}}\int_{0}^{t_{f}}{\rm d}t\ \theta_{Q}\left(t\right)\sin\left(\omega_{n}t\right).
		\end{align}
		Combining these we obtain
		\begin{align}
		& \exp{\left(-\frac{\beta}{2}\left(\omega_{n}^{2}A^{2}+B^{2}\right)\right)} =  \exp\left(-\frac{c_{n}^{2}}{2\omega_{n}}k_{B}T\int^{t_{f}}{\rm d}t\int^{t_{f}}{\rm d}t^{\prime}\ \theta_{Q}\left(t\right)\gamma_{n}\left(t-t^{\prime}\right)\theta_{Q}\left(t^{\prime}\right)\right) 
		\end{align} 
		using $\gamma_{n}\left(t-t^{\prime}\right) =\frac{1}{\omega_{n}}\cos\left(t-t^{\prime}\right)$. Collecting these results, we are able to express the influence functional
		\begin{align}
		\mathcal{F}_{\rm cl}=&{\rm e}^{-\sum_{n}\frac{c_{n}^{2}}{2\omega_{n}}\Phi_{n}} ,
		\end{align}
in terms of the influence phase
		\begin{align}
		\Phi_{n}=&-2i\int_{0}^{t_{f}}{\rm d}t\ \theta_{Q}\left(t\right)\int_{0}^{t}\textrm{d}t^{\prime}\,q(t^{\prime})\sin\left(\omega_{n}(t-t^{\prime})\right) \nonumber \\ &+k_{B}T\int^{t_{f}}{\rm d}t\int^{t_{f}}{\rm d}t^{\prime}\ \theta_{Q}\left(t\right)\gamma_{n}\left(t-t^{\prime}\right)\theta_{Q}\left(t^{\prime}\right).
		\end{align}
		
		At this point we take the continuum limit for the oscillators, 
		\begin{align}
		\sum_{n}\frac{c_{n}^{2}}{2\omega_{n}}\to\int_{0}^{\infty}\frac{{\rm d}\omega}{2\pi}\ I\left(\omega\right) \label{eq:spectrumdimensions}
		\end{align} 
		such that our final influence functional is given by
		\begin{align}
		\mathcal{F}_{\rm cl} =&\exp \left[2i\int_{0}^{t_{f}}{\rm d}t\ \theta_{Q}\left(t\right)\int_{0}^{t}\textrm{d}t^{\prime}\,q(t^{\prime})\frac{{\rm d}\gamma\left(t-t^{\prime}\right)}{{\rm d}t^{\prime}}\right. \nonumber \\ &\left.-\int^{t_{f}}{\rm d}t\int^{t_{f}}{\rm d}t^{\prime}\ \theta_{Q}\left(t\right)k_{B}T\gamma\left(t-t^{\prime}\right)\theta_{Q}\left(t^{\prime}\right)\right],
		\label{eq:preHSclassicalinfl} \\
	\gamma\left(t-t^{\prime}\right)=&\int_{0}^{\infty}\frac{{\rm d}\omega}{\omega\pi}\ I\left(\omega\right)\cos\left(t-t^{\prime}\right). \label{eq:gamma}
	\end{align}

Having evaluated the influence functional, several possibilities now present themselves. The overall path integral for the propagator is quadratic in the $\theta_Q$ variable, and it is therefore possible to integrate the entire path integral over  $\theta_Q$ to leave a path integral only in the $q$ variable. To perform this integral, we would require the inverse kernel $\gamma^{-1}(t)$ \cite{doi:10.1142/2113} which satisfies
\begin{equation}
\int^t_0{\rm d}\tau^\prime \gamma(\tau-\tau^\prime)\gamma^{-1}(\tau^\prime-\tau^{\prime \prime})=\delta(\tau-\tau^{\prime \prime}).
\end{equation}

Returning to the discrete form and performing this integration would leave us with a path integral where path weightings are determined by the \emph{Onsager-Machlup} functional \cite{Onsager1953,Hanggi1989}, as shown in Fig.\ref{fig:influenceshematic}. Rather than pursuing this method, which would simply replace one path integral with another, we would instead like to work backwards from the path integral to obtain an effective equation of motion.  While the first term in the exponent of $\mathcal{F}_{\rm cl}$  is of the required form, the second term is quadratic with respect to the $\theta_Q$, and we therefore require a mapping that brings this term to the form required by Eq. (\ref{influenceforeqnmotion}) to construct an effective equation of motion. Specifically, we seek a function $\eta(t)$ satisfying: 
\begin{equation}
i\int_{0}^{t_{f}}{\rm d}t\ \theta_{Q}\left(t\right) \eta(t) = \kappa(t_f) \label{eq:Volterra},
\end{equation}
where $\kappa(t_f)$ is defined as
\begin{equation}
\kappa(t_f)=-\int^{t_{f}}{\rm d}t\int^{t_{f}}{\rm d}t^{\prime}\ \theta_{Q}\left(t\right)k_{B}T\gamma\left(t-t^{\prime}\right)\theta_{Q}\left(t^{\prime}\right).
\end{equation}
Here Eq.(\ref{eq:Volterra}) is a Volterra equation of the first kind with respect to $\eta$ \cite{polyanin2008handbook}, with the unique solution
\begin{equation}
    \eta(t)=-i\frac{\dot{\kappa}(t)}{\theta_Q(t)}= 2ik_{B}T\int^t_0{\rm d}t^{\prime}\ \gamma\left(t-t^{\prime}\right)\theta_{Q}\left(t^{\prime}\right).
\end{equation}
Clearly, this solution is functionally dependent on $\theta_Q$, and therefore not admissible as part of an effective equation of motion. From this we conclude that an influence functional of the form of form required by Eq.(\ref{influenceforeqnmotion}) is unattainable when $\eta$ is a deterministic function of $t$. 

In order to remedy this problem, and generate an effective equation of motion, we consider the possibility that ${\rm e}^{\kappa(t)}$ is the average of a stochastic term. Specifically, we interpret it as the \emph{characteristic function} of a probabilistic process $W(\eta(t))$ describing a random variable $\eta(t)$:
\begin{equation}
{\rm e}^{\kappa(t_f)}= \left\langle {\rm e}^{i\int_{0}^{t_{f}}\mathrm{d}t\ \eta(t)\theta_Q(t)}\right\rangle _{r}, \label{characteristicfunction}
\end{equation}
where the statistical average is given by
\begin{equation}
\left\langle {\rm e}^{i\int_{0}^{t_{f}}\mathrm{d}t\  \eta(t)\theta_Q(t)}\right\rangle _{r}= \int \mathcal{D}\eta(t) \ W(\eta(t))  {\rm e}^{i\int_{0}^{t_{f}}\mathrm{d}t\ \eta(t)\theta_Q(t)}.
\end{equation} 
The characteristic function \emph{uniquely} determines the process $W$ and hence the statistical properties of $\eta(t)$. It therefore follows that if a process can be found whose characteristic function is ${\rm e}^{\kappa(t)}$, we may replace the deterministic $\kappa$ term (which is quadratic in $\theta_Q$) with a stochastic one linear in $\theta_Q$, on the understanding that the true effective propagator must be averaged over the process $W$. This would then satisfy the form given in Eq.(\ref{influenceforeqnmotion}), which is necessary to construct an effective equation of motion.  
\begin{figure}
\centering
\includegraphics[height=5cm]{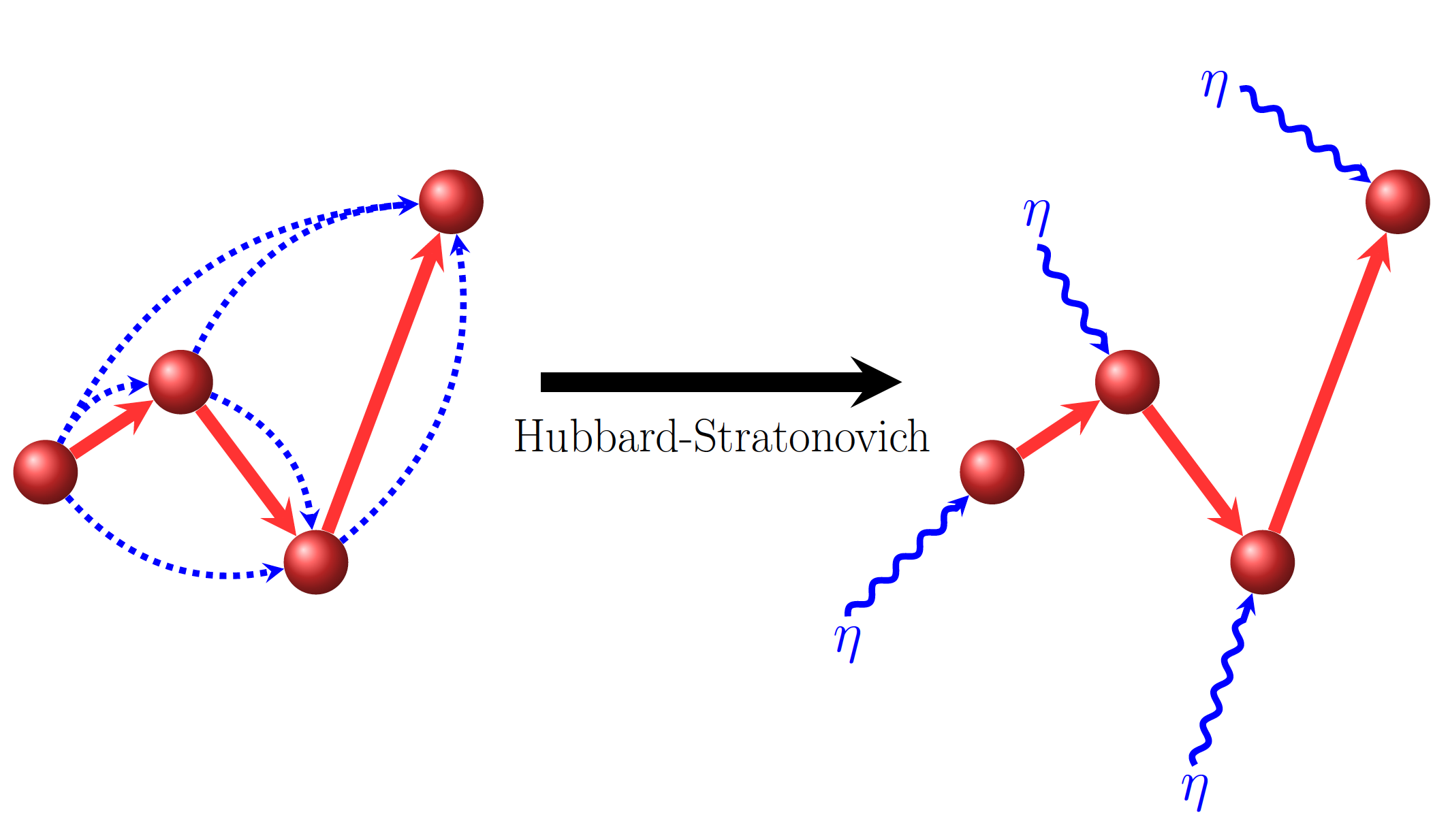} \caption{On the left hand side the system evolution is influenced by earlier behaviour (indicated by dashed blue lines), which by a Hubbard-Stratonovich transformation is equivalent to the right hand side, showing local dynamics with an additional stochastic term $\eta$. \label{fig:HStransform}}
\end{figure}

To find the process $W$, we employ the \emph{Hubbard-Stratonovich} (HS) transformation, which equates a deterministic non-local integral exponent
to one involving local stochastic terms that must be averaged over
a distribution $W$. Explicitly, the HS transform states 
\begin{align}
	\left\langle {\rm e}^{i\int_{0}^{t_{f}}\mathrm{d}t\  z(t)k(t)}\right\rangle _{r} = {\rm e}^{-\frac{1}{2} \int_{0}^{t_{f}}\int_{0}^{t_{f}}\mathrm{d}t\textrm{d}t^{\prime}\ k(t)\Sigma (t,t^\prime)k(t^{\prime})}, \label{eq:HSindentitymaintext}
	\end{align} 
where $z(t)$ is a zero-mean Gaussian random variable with correlation function $\langle z(t) z(t^\prime) \rangle=\Sigma(t,t^\prime)$. A full derivation of this transformation may be found in appendix \ref{HStransform}, but here it is sufficient to emphasise that the HS transform is a formally exact procedure, which uniquely defines the statistical properties of $z(t)$. Specifically, the kernel of the right-hand two body term defines the correlation function of the one-body stochastic process that it is being mapped to. In a more physical sense, we can interpret the
HS transformation as converting a system of two body potentials into
a set of independent particles in a fluctuating field. Fig. \ref{fig:HStransform} sketches the physical equivalence via the HS transform between a system where the motion of the particle is influenced by its earlier behaviour, and one in which the two-body potential is replaced by a stochastic term. 

To apply this to the CIF, we note that the right hand side of the HS transform is precisely  ${\rm e}^{\kappa(t)}$ using the substitutions $k(t)\to -\theta_Q (t)$ and $\Sigma(t,t^\prime) \to 2k_BT \gamma(t-t^\prime)$. We may therefore re-express the problematic term in the CIF using the HS transformation:
	\begin{align}
	\exp\left(-\frac{1}{2}\int^{t_{f}}{\rm d}t\int^{t_{f}}{\rm d}t^{\prime}\ \theta_{Q}\left(t\right)2k_{B}T\gamma\left(t-t^{\prime}\right)\theta_{Q}\left(t^{\prime}\right)\right)\nonumber  \\ =\left\langle \exp\left(-i\int_{0}^{t_{f}}{\rm d}t\ \eta_{{\rm cl}}\left(t\right) \theta_{Q}\left(t\right)\right)\right\rangle _{r},
	\end{align}
with the stochastic term defined by its moments
	\begin{align} \left\langle {\rm \eta}_{{\rm cl}}\left(t\right)\right\rangle _{r} =&0, \nonumber \\
	 \left\langle {\rm \eta}_{{\rm cl}}\left(t\right){\rm \eta}_{{\rm cl}}\left(t^{\prime}\right)\right\rangle _{r}=&2k_{B}T\gamma\left(t-t^{\prime}\right).
	\end{align}
	 Thus, using the formally exact HS transformation we have obtained a uniquely defined noise term, a powerful equivalence that is only available using the path integral formulation. Furthermore, this method establishes the necessity of stochasticity by demonstrating the corollary statement that there is no deterministic equation of motion that corresponds to the reduced system dynamics. 
	
	Putting all of this together, we find that the KvN propagator is given by the averaging over the propagators for a single stochastic trajectory, $U_{\rm cl}=\langle{\tilde{U}_{{\rm cl}}}\rangle_r$, where
	\begin{align}
	\tilde{U}_{{\rm cl}}=&\int_{q_{0},\dot{q}_{0}}^{q_{f},\dot{q}_{f}}\mathcal{D}q(t)\mathcal{D}\theta\left(t\right)\ {\rm e}^{i\int_{0}^{t_{f}}{\rm d}t\ \theta\left(t\right)R\left(t\right)}, \\ R\left(t\right)=&m\ddot{q}\left(t\right)+V^{\prime}\left(q,t\right)-\eta_{{\rm cl}}\left(t\right)-2\int_{0}^{t}\textrm{d}t^{\prime}\,q(t^{\prime})\frac{{\rm d}\gamma\left(t-t^{\prime}\right)}{{\rm d}t^{\prime}}.
	\end{align}

The final term in $R$ may be integrated by parts:
	\begin{align}
	\int_{0}^{t}\textrm{d}t^{\prime}\,q(t^{\prime})\frac{{\rm d}\gamma\left(t-t^{\prime}\right)}{{\rm d}t^{\prime}}=&q\left(t\right)\gamma\left(0\right)-q\left(0\right)\gamma\left(t\right) -\int_{0}^{t}\textrm{d}t^{\prime}\,\dot{q}(t^{\prime})\gamma\left(t-t^{\prime}\right).
	\end{align}
The first two terms are pure functions of time of $q$, and hence can be absorbed into the arbitrary potential $V$, leaving only the friction term. If we had included the interaction in our original thermal density, we would have had an extra term cancelling $q\left(0\right)\gamma\left(t\right)$ here, while including the counterterm in the open system Hamiltonian would cancel $q\left(t\right)\gamma\left(0\right)$. Substituting this back into the propagator and performing the path integral over $\theta\left(t\right)$ we obtain:
	\begin{align}
	\tilde{U}_{{\rm cl}} =&\int_{q_{0},\dot{q}_{0}}^{q_{f},\dot{q}_{f}}\mathcal{D}q(t)\ \delta\bigg[m\ddot{q}\left(t\right)+V^{\prime}\left(q,t\right)\bigg. \left. +2\int_{0}^{t}\textrm{d}t^{\prime}\,\dot{q}(t^{\prime})\gamma\left(t-t^{\prime}\right)-\eta_{{\rm cl}}\left(t\right)\right] \label{eq:Langevinpropagator}
	\end{align}
	
	This brings us to the ultimate result of this section, namely that the equation of motion for a single trajectory is a generalised Langevin equation:
	\begin{align}
	m\ddot{q}\left(t\right)=-V^{\prime}\left(q,t\right)-2\int_{0}^{t}\textrm{d}t^{\prime}\,\dot{q}(t^{\prime})\gamma\left(t-t^{\prime}\right)+\eta_{{\rm cl}}\left(t\right). \label{eq:Langevinwithfriction}
	\end{align} 
	In the particular case where $I\left(\omega\right)=D\omega$, we recover a Markovian Langevin equation, with $\left\langle \eta_{{\rm cl}}\left(t\right)\eta_{{\rm cl}}\left(t^{\prime}\right)\right\rangle _{r}=2k_{B}TD\delta\left(t-t^{\prime}\right)$, $\gamma\left(t\right)=D\delta\left(t\right)$.
	
\begin{figure}
\centering
\includegraphics[height=5cm]{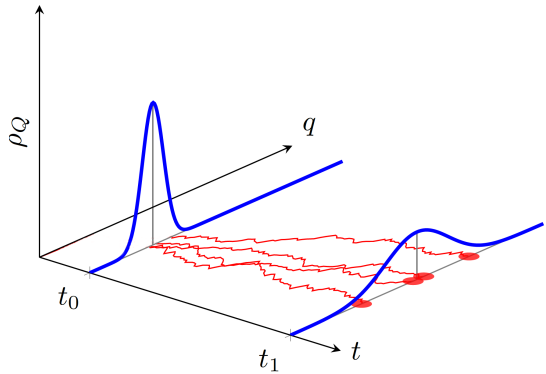} \caption{The HS transform maps the effective propagator to the average of a set of stochastic propagations. By averaging over the trajectories these propagators generate, one is able to obtain the true reduced probability distribution evolved by the effective propagator. \label{fig:stochastictrajectories}}
\end{figure}

In order to obtain the reduced probability density, we must average over the stochastic propagations of the initial density:
\begin{equation}
    \rho_Q(q_f,p_f)=\left<\tilde{U}_{\rm cl} \rho_Q (q_0,p_0) \right>_r.
\end{equation}
This effectively corresponds to constructing the distribution from the average number of trajectories that end at each point in the phase space. Figure \ref{fig:stochastictrajectories} illustrates some sample trajectories, together with the probability distribution their average describes.

\subsection{Quantum Case: The Stochastic Liouville-von Neumann Equation}
We now turn our attention to the derivation of an effective equation for quantum dynamics. The process is in principle identical to the classical case, with an important caveat regarding initial conditions. 
Unlike in the classical case, neglecting the system-environment interaction in the initial state is only appropriate when this interaction is weak. At strong coupling, it is necessary to account for the true initial equilibrium $\hat{\rho}_{0}^{{\rm tot}}=\frac{1}{Z_\beta}e^{-\beta\hat{H}_{\rm tot}}$. This generalisation can be accounted for in a number of ways \cite{Ankerhold2005}, with one option being to include it as an imaginary time path integral within the influence functional \cite{Grabert1988, Ankerhold2001, Stockburger2002,ourpaper}. For the sake of simplicity however we will again adopt a \emph{partitioned} initial condition:
\begin{equation}
    \hat{\rho}_{0}^{{\rm tot}}=\hat{\rho}_{Q}\otimes\hat{\rho}_{\beta}
\end{equation}
where the initial environment density matrix is the thermal matrix for the non-interacting oscillators:
\begin{equation}
    \hat{\rho}_{\beta}=\prod_n \frac{{\rm e}^{-\frac{1}{2}\beta\left(\hat{k}^2_n+\omega^2_n\hat{x}^2_n\right)}}{2\sinh\left(\frac{1}{2}\omega_n\hbar\beta\right)}.
\end{equation}

With this setup, one follows the same procedure as the classical case, formulating and evaluating the environmental part of the path integral. Once again, this amounts to solving a number of straightforward but tedious Gaussian integrals. A full derivation of the influence functional for the CL model can be found in a number of sources\cite{Kleinertbook,KLEINERT1995, Grabert1988, ourpaper}, and after introducing sum-difference coordinates 
\begin{equation}
\epsilon\left(t\right)=q\left(t\right)-q^{\prime}\left(t\right), \quad y\left(t\right)=\frac{1}{2}\left(q\left(t\right)+q^{\prime}\left(t\right)\right),
\end{equation}
the final influence phase reads \cite{Grabert1988, ourpaper}
 \begin{align}
\mathcal{F}\left[q\left(t\right),q'\left(t\right)\right]={\rm e}^{\frac{i}{\hbar}\Phi\left[q\left(t\right),q'\left(t\right)\right]}, 
\end{align}
\begin{align}
\Phi\left[q,q^{\prime}\right]=&\frac{1}{2}\int_{0}^{t_{f}}\mathrm{d}t\int_{0}^{t_{f}}\mathrm{d}t^{\prime}\,K^{R}\left(t-t^{\prime}\right)\epsilon\left(t\right)\epsilon\left(t^{\prime}\right)+2i\int_{0}^{t_{f}}\mathrm{d}t\int_{0}^{t}\mathrm{d}t^{\prime}\,K^{I}\left(t-t^{\prime}\right)\epsilon\left(t\right)y\left(t^{\prime}\right) \label{eq:qinfluencephase}.
\end{align}
Here the kernels are defined as
\begin{align}
K^{R}\left(t-t^{\prime}\right)=&\hbar\int_{0}^{\infty}\frac{{\rm d}\omega}{\pi}\,I\left(\omega\right)\coth\left(\frac{1}{2}\omega\hbar\beta\right)\cos\left(\omega\left(t-t^{\prime}\right)\right), \label{eq:Corr_F_Eta_Etasimple2} \\
K^{I}\left(t\right)=&\frac{{\rm d}\gamma\left(t-t^{\prime}\right)}{{\rm d}t},
\end{align}
where $\gamma(t)$ is unchanged from Eq.\eqref{eq:gamma}. Note that $\Phi$ must have the dimensions of action, which in turn sets the dimensions of both kernels and their constituent components, i.e. both $I(\omega)$ and $\gamma(t)$ have dimension $T^{-2}$, as can also been seen from Eqs.(\ref{eq:spectrumdimensions},\ref{eq:gamma}).

This structure is strikingly similar to the classical influence, and we encounter the same obstacle to constructing an equation of motion, namely that there are terms which prevent the effective propagator being cast in the form of Eq.\eqref{eq:quantumeffpropagatordisentangle}. Naturally, the solution is once again to employ the HS transformation. In this case, we apply it to both the $\epsilon(t)$ and $y(t)$ coordinates, such that the influence functional may now be described by
\begin{equation}
    \mathcal{F}\left[q\left(t\right),q'\left(t\right)\right]=\left\langle\exp\left(\frac{i}{\hbar}\int^{t_f}_0 \eta(t)\epsilon(t) +\nu(t)y(t)\  {\rm d}t\right)\right\rangle_r
\end{equation}
with the two \emph{complex-valued} noises $\eta(t)$ and $\nu(t)$ \cite{Stockburger2002} defined by the correlation functions: 
\begin{align}
\left\langle \eta\left(t\right)\eta\left(t^{\prime}\right)\right\rangle _{r}&=K^R\left(t-t^{\prime}\right),\label{eq:Corr_F_Eta_Etasimple} \\
\left\langle \eta(t)\nu\left(t^{\prime}\right)\right\rangle _{r}&=-2i\hbar \Theta\left(t-t^{\prime}\right)K^{I}\left(t\right) \label{eq:Corr_F_Eta_Nusimple} \\
\left\langle \nu(t)\nu\left(t^{\prime}\right)\right\rangle _{r}&=0.
\end{align}
Importantly, the HS transformation allows some freedom in the definition of the noises, and that in this instance while $K^R$ and $K^I$ have the same dimension, the physical dimension of $\nu$ has been chosen with an additional dimension of action relative to the $\eta$ noise, which accounts for the factor of $\hbar$ in Eq.\eqref{eq:Corr_F_Eta_Nusimple}.

From this transformation it is now possible to define effective forward and backward propagators (in the original $q, q^\prime$ coordinates) for a given stochastic realisation:
\begin{equation}
\hat{U}^{\pm}\left(t_{f}\right)=\widehat{T}^\pm\exp\left(\mp \frac{i}{\hbar}\int_{0}^{t_{f}}\left(\hat{H}_{Q}\left(t\right)-\left[\eta\left(t\right)\pm\frac{1}{2}\nu\left(t\right)\right]\hat{q}\ \right)\mbox{d}t\right)\label{eq:classprop}.
\end{equation}
Such that the single-trajectory density matrix $\tilde{\rho}(t)$ evolves in the following manner
\begin{equation}
    \tilde{\rho}(t)=\hat{U}^+(t) \tilde{\rho}(0)\hat{U}^-(t),
\end{equation}
where ($\hat{T}^-$ ) $\hat{T}^+$ is the (anti) time ordering operator. From here it is trivial to derive the ultimate result of this section, the \emph{Stochastic Liouville-von Neumann Equation} (SLE) \cite{Stockburger2004} :
\begin{align}
	i\hbar\frac{{\rm d}\tilde{\rho}\left(t\right)}{{\rm d}t}=&\left[\hat{H}_{Q}\left(t\right),\tilde{\rho}\left(t\right)\right]-\eta\left(t\right)\left[\hat{q},\tilde{\rho}\left(t\right)\right]+\frac{1}{2}\nu\left(t\right)\left\{ \hat{q},\tilde{\rho}\left(t\right)\right\}. \label{eq:SLEstockburger}
	\end{align}
The SLE evolves a single-trajectory density matrix  $\tilde{\rho}(t)$ which upon stochastic averaging gives the physical reduced density matrix
\begin{equation}
\left\langle \tilde{\rho}(t)\right\rangle_r=\hat{\rho}_Q(t).
\end{equation}
Note that in some circumstances one is also able to perform the stochastic averaging at the level of the equation of motion itself, in order to recover the standard master equation representation for open system dynamics \cite{Yan2016}.

The relationship between the stochastic evolution of a single-trajectory density matrix, and the physical density of the open system is of critical importance, as it is from this one may calculate useful expectations of the open system. It does however beg the question as to whether an individual trajectory can be assigned a physical interpretation, which we now address.

\subsection{Interpreting Trajectories}
In both the classical and quantum cases, we have found that for an archetypal environment model, the equations of motion become stochastic. The stochastic trajectories one must average over arose as a formal device in the HS transformation
for describing an effective Hamiltonian, without reference to the
physical content of this transformation. The interpretation of stochastic
terms in the classical context is straightforward \cite{Kleinertbook},
but what about in the fully quantum case? Is it possible to attach
a physical interpretation to an individual trajectory? 

Critically, for an element of a formalism to be physically meaningful,
it must be possible to isolate its effect on observations. In the
purely classical case this is not a problem, as any given trajectory
has an associated probability to be observed, entirely dependent on
the initial condition of the combined system and bath. The quantum case is no different, in the sense that the stochastic
terms are also capturing the effect of an unknown initial state sampled
from some probability distribution. 

The difference between the classical and quantum cases lies in the fact
that in the quantum regime the statistical distribution of an observable
is due to both the probabilistic sampling of an initial state, \emph{and
}the inherently quantum nature of the system evolution. In order to
draw out a physical interpretation for a single stochastic trajectory, it must be possible to disentangle
these two contributions, such that one can identify the ensemble of
realisations generated by the \emph{same} initial state. 

\begin{figure}
\begin{centering}
\includegraphics[width=1\textwidth]{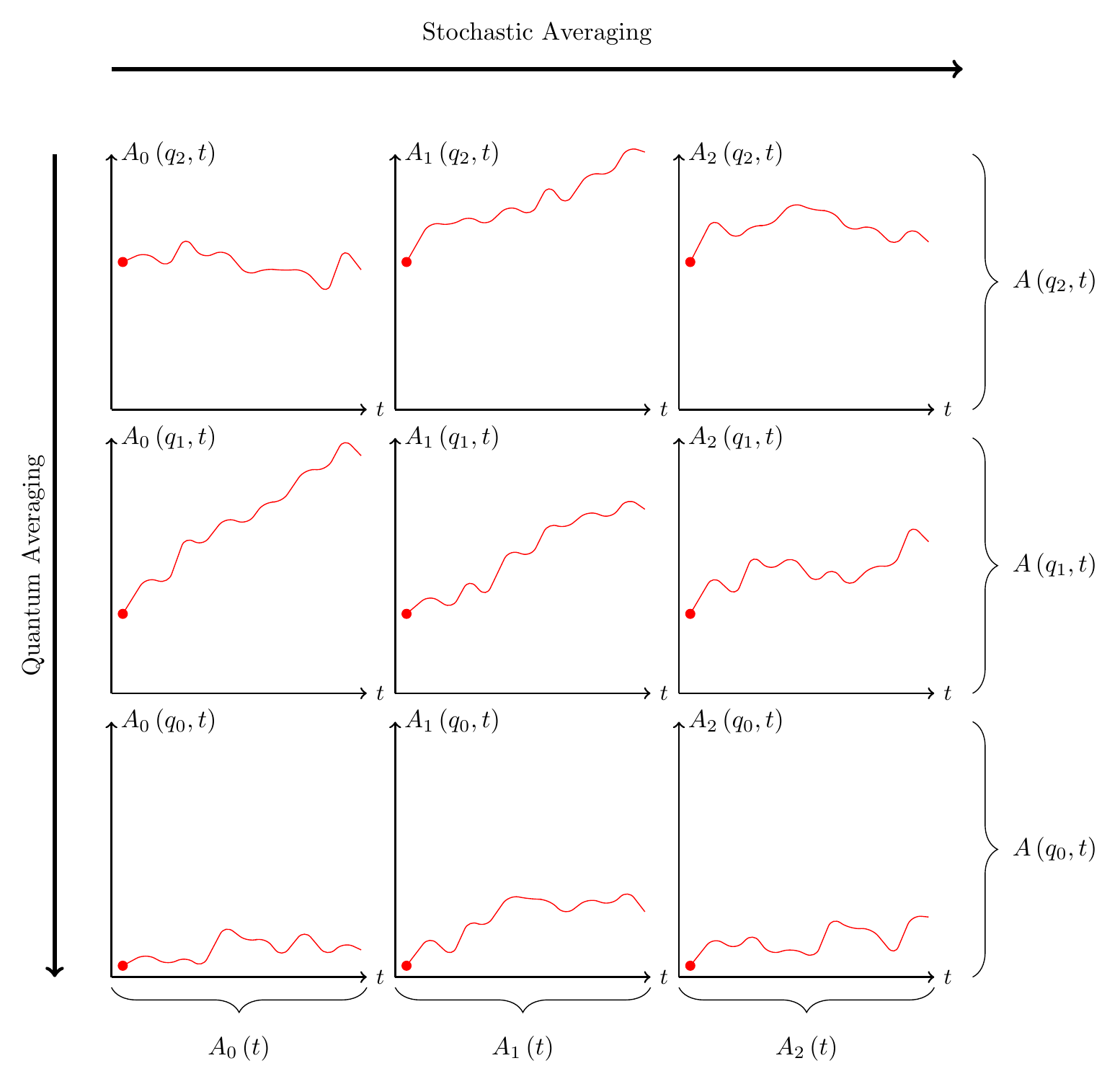} 
\par\end{centering}
\caption{Schematic demonstrating the full ensemble over different initial system+bath
states. Each trajectory represents a single measurement outcome (labelled by $q$) when evolving from some initial environment state (labelled by the subscript $j$). The braces indicate the result when averaging over the appropriate
dimension. Averaging over both individual measurement outcomes and stochastic trajectory
would yield the experimentally observed expectation $\left\langle \hat{A}\left(t\right)\right\rangle $=$A\left(t\right)$.
\label{fig:ensembleplusstochastic}}
\end{figure}

Explicitly, let us consider an observable expectation $A$ for the
$Q$ subsystem. The expectation of such an observable will be given
by
\begin{equation}
A(t)=\left\langle \hat{A}\left(t\right)\right\rangle ={\rm Tr}\left(\left\langle \tilde{\rho}\left(t\right)\right\rangle _{r}\hat{A}\right)=\int{\rm d}q\ \rho(q,q;t)A\left(q,t\right). \label{eq:Aaverage}
\end{equation}
As both the trace operation and the stochastic averaging are linear
operations, we can swap the order of averaging
\begin{equation}
A(t) =\left\langle {\rm Tr}\left(\tilde{\rho}\left(t\right)\hat{A}\right)\right\rangle _{r}=\left\langle A_{j}\left(t\right)\right\rangle _{r}.
\end{equation}
The object $A_{j}\left(t\right)$ is the quantum average for a single realisation of the noise, labeled by $j$. To make this concrete, evolving with any given realisation of $\tilde{\rho}$ will be equivalent to evolving the system when the environment is initially in some pure state. For the sake of simplicity, let us assume that the $Q$ system is initially in some known pure state $\ket{\psi_Q}$, such that any particular initial state may be given by:
\begin{equation}
    \ket{\psi_j}=\ket{\psi_Q}\otimes\ket{E_j}
\end{equation}
where  $\ket{E_j}$ is some unknown energy eigenstate of the environment, drawn with probability $\frac{{\rm e}^{-\beta E_j}}{Z_\beta}$. In this picture, our expectation will be given by 
\begin{equation}
A_j(t)= \left\langle\psi_j\left|\hat{U}_{\rm tot}^\dag(t)(\hat{A}\otimes\mathbbm{1}_X)\hat{U}_{\rm tot}(t)\right|\psi_j\right\rangle.
\end{equation}
Naturally, this quantum expectation can be expressed as an averaging over different measurement outcomes. For the sake of consistency with Eq.\eqref{eq:Aaverage}, we will express this expectation in the coordinate basis, inserting resolutions of unity $\int {\rm d}q{\rm d}\vec{x}\ \ket{q,\vec{x}}\bra{q,\vec{x}}$ as appropriate to obtain
\begin{align}
A_{j}\left(t\right)&=\int{\rm d}q\ p_j(q,t) A_{j}\left(q,t\right) \\
p_j(q,t)&=\int{\rm d}\vec{x}\  \left|\left\langle q,\vec{x}\left|\hat{U}_{\rm tot}(t)\right|\psi_j\right\rangle\right|^2.
\end{align}
Here $p_j$ quantifies the probability of observing a particular measurement outcome, and $A_{j}\left(q,t\right)$ is the value of that measurement starting from some pure state $\ket{\psi_j}$. In this way, we can think of the final expectation $A(t)$ as consisting of both a quantum averaging over measurement outcomes, and a stochastic averaging over initial environment pure states. The integration of $A_{j}\left(q,t\right)$ over its $q$
coordinate performs the quantum average, while averaging over the $j$
index performs the stochastic average. The two different processes are illustrated in Fig. \ref{fig:ensembleplusstochastic},
demonstrating the way one may partition the expectations associated with each type of averaging, with the full
observable expectation being recovered when averaging over both.  In
this sense it is possible to assign an interpretation to a given stochastic
trajectory, i.e. that its expectations are what we would observe for
a system evolving from a given pure state for the composite system. 

Of course, for any single measurement process, our initial state is drawn from a distribution. In order to perform the quantum averaging, it must be possible to collect statistics on this unknown pure initial state. To do so, this unknown state must be copyable onto a different system in some state  $\ket{T}$. The question of whether this copying is possible is crucial to separating the stochastic effects determined by the initial condition, and the quantum indeterminacy baked into the formalism. If one were able to clone a given system+environment initial state many times and perform averages on its evolution, then $A_{j}\left(t\right)$ would be
directly observable and undeniably physical.   

Clearly then, the interpretibility of stochastic trajectories turns on whether there is a unitary operation  $\hat{C}$ that can operate on all possible initial states such that
\begin{equation}
    \forall \ket{\psi_j}, \hat{C}\left( \ket{\psi_j}\otimes\ket{T}\right)=\ket{\psi_j}\otimes\ket{\psi_j}.
\end{equation}
Note that simply from linearity, $\hat{C}$ can only clone basis states. If for example, one applied the cloning operation to $\ket{\phi}= \frac{1}{\sqrt{2}} \left(\ket{\psi_j} +\ket{\psi_k}\right)$, then the result would be:
\begin{equation}
    \hat{C}\left(\ket{\phi}\otimes\ket{T}\right)= \frac{1}{\sqrt{2}} \left(\ket{\psi_j}\otimes\ket{\psi_j}+\ket{\psi_k}\otimes\ket{\psi_k}\right)\neq \ket{\phi}\otimes\ket{\phi}. \label{eq:clonesuperposition}
\end{equation}
This immediately implies that for our cloning operation to work, there is some basis where every possible initial environment state can be expressed as a pure state. Let us presume for the moment that this is true, and assume that $\hat{C}$ exists for this basis. Now consider two states $\ket{\psi_j}$ and $\ket{\psi_k}$. Performing the copying operation on each, and taking the inner product we have
\begin{equation}
   \braket{\psi_j|\psi_k}\braket{T|T}= \left(\bra{\psi_j}\otimes\bra{T}\right)\hat{C}^\dag\hat{C}\left( \ket{\psi_k}\otimes\ket{T}\right)= \left|\braket{\psi_j|\psi_k}\right|^2 \label{eq:cloneorthogonality}
\end{equation}
from which we conclude that any state we wish to copy must satisfy $\braket{\psi_j|\psi_k}=\left|\braket{\psi_j|\psi_k}\right|^2$, i.e. $\braket{\psi_j|\psi_k}=\delta_{jk}$. This is problematic, as any environmental system is very likely to be highly degenerate. Of course, any degenerate basis can always mapped to an orthogonal basis. If (for example) $\ket{\psi_j}$ and $\ket{\psi_k}$ are degenerate, one could perform an orthogonalisation procedure, defining new basis states $\ket{\psi_\pm}= \frac{1}{\sqrt{2}} \left(\ket{\psi_j} \pm \ket{\psi_k}\right)$. Note however that regardless of basis chosen, all of $\ket{\psi_j}$ and $\ket{\psi_k}$ and $\ket{\psi_\pm}$  are valid initial states for a realisation of the system but belong to different bases. Therefore by Eqs.(\ref{eq:clonesuperposition},\ref{eq:cloneorthogonality}) there is no unitary procedure that can successfully clone all of these states simultaneously. 

Without knowing the state we are trying to clone, there is no way of knowing the appropriate basis that the cloning procedure $\hat{C}$ should work in. As a result, one cannot copy a given realisation of the combined system and environment initial state. This problem is an example of the quantum no-cloning theorem \cite{Wootters1982}, which explicitly prohibits the copying
of an arbitrary state. In this case, the degeneracy of the eigenstates of the bath forbids any procedure that can take an unknown eigenstate and copy it. This result can be understood intuitively in the sense that one could perform a measurement on the energy of the bath before the evolution begins,  but this only uniquely determines the initial state of the system when all energy eigenvalues of the bath are distinct. If the initial state can't be uniquely identified, then isolating its effect of that particular state on the evolution of the system is similarly impossible. 

Naturally, there are loopholes to this argument - recent work has shown there are \emph{approximate} theorems where there is some probability of cloning an arbitrary state successfully \cite{Rui2018}. In this case however the problem is just transposed to knowing if the cloning of an unknown state has been successful or not, and does not materially affect the argument presented here. More generally, if the restrictions on $\hat{C}$ being unitary (and therefore linear) are relaxed, then there may be some way to clone an unknown state with perfect fidelity, but it is unclear how one would construct such an operator. For this reason, under reasonable assumptions, $A_j (t)$ becomes impossible to access experimentally. Excepting the edge case of a non-degenerate bath spectrum, an arbitrary initial state responsible for a single stochastic realisation cannot be copied, and therefore the SLE only acquires meaning after performing the stochastic
averaging over these non-copyable initial states. 

Having established this result, it is important to distinguish between the stochasticity derived from the influence functional including an integration over initial environment states, and the stochasticity induced by a continuous measurement process \cite{doi:10.1080/00107510601101934,jacobs2014quantum}. In the latter case, the stochasticity is of a qualitatively different origin, and one is able to directly observe (by definition) a single trajectory generated from constant measurement. 

Finally, we note that this uninterpretability of individual trajectories derived from influence functionals is a purely quantum phenomenon, despite the fact that even in the classical limit there is an equivalent no-cloning theorem \cite{PhysRevLett.88.210601}. In the classical case, the Hilbert space averaging process
becomes redundant (as observables commute), meaning a single trajectory can be associated with an initial state without the need to average over unrealisable copies of that initial state. Expressed differently, in the classical case the environment Hamiltonian can be always be represented in the phase-space basis, which is a non-degenerate set of eigenstates for all bath observables. Hence, an unproblematic interpretation of the trajectories
is recovered.

\section{The CIF as a Classical Limit\label{classicallimit}}
As a final demonstration of the utility of influence functionals,  we use them to establish that classical limit of the SLE seen in the previous section is indeed the generalised Langevin equation. Taking the classical limit of this (and indeed any) system is by no means straightforward, as the classical limit is itself ill-defined (see Ref.\cite{Cabrera2019}, which discusses this in detail). Here we shall follow the example of Ref.\cite{Cabrera2019}, defining the classical limit as that of a commutative algebra between observable operators. This can be achieved in the usual way by taking $\hbar \to 0$. In the case of the SLE there is the additional complication that the stochastic term correlations \emph{also} involve a factor of $\hbar$ which must also be limited to zero. As we shall see, difficulties arise from this which require the CIF to resolve.

\subsection{Heuristic Limit}
To understand some of the difficulties associated with taking the classical limit of the SLE, we first make a heuristic calculation, noting that in this limit, the only path that contributes to the propagator is that with the action $\tilde{S}_{{\rm cl}}^{\pm}$
\begin{equation}
\tilde{S}_{{\rm cl}}^{\pm}=\int_{0}^{t_{f}}\mbox{d}t\ \left[ L_{Q}\left(q_{{\rm cl}}(t)\right) + \left[\eta_{{\rm cl}}\left(t\right)\pm\frac{1}{2}\nu_{{\rm cl}}\left(t\right)\right]q_{{\rm cl}}\left(t\right) \right], \label{eq:Sclassicsimple}
\end{equation}
where $ L_{Q}\left(q_{{\rm cl}}(t)\right)$ indicates the Lagrangian for the system following the classical equation of motion. The correlation functions for the noises will also be affected in
the classical limit, hence we define new noises that obey these limiting
correlation functions:
\begin{align}
\lim_{\hbar\to0}\left\langle \eta\left(t\right)\eta\left(t^{\prime}\right)\right\rangle _{r}&= \left\langle \eta_{{\rm cl}}\left(t\right)\eta_{{\rm cl}}\left(t^{\prime}\right)\right\rangle _{r}=2k_{B}T \gamma \left(t-t^\prime \right), \\
\lim_{\hbar\to0}\left\langle \eta\left(t\right)\nu\left(t^{\prime}\right)\right\rangle _{r}&=0.
\end{align}

The statistics of the $\eta$ noise becomes identical to that derived previously, and since the $\nu$ noise is now entirely uncorrelated, it will
have no effect on the average dynamics and can be dropped from the
action. This restores symmetry to the forwards and backwards propagations
\begin{equation}
\tilde{S}_{{\rm cl}}=\int_{0}^{t_{f}}\mbox{d}t\ \left[ L_{Q}\left(q(t)\right)+\eta_{\rm cl}\left(t\right)q\left(t\right) \right].
\end{equation}
The classical equation of motion we obtain for a single trajectory
is therefore a type of Langevin equation:
\begin{equation}
\ddot{q}=-\frac{\partial V\left(q\right)}{\partial q}-\eta_{{\rm cl}}\left(t\right). \label{eq:classicallimitfirst}
\end{equation}

It is not a surprise that the classical limit of the SLE corresponds
to a Langevin equation, but Eq. (\ref{eq:classicallimitfirst}) appears
to lack the essential feature of friction, as the $\nu$
noise appears to have no effect on the dynamics. This is a consequence
of incorporating the dynamic response of the bath into the HS transformation,
and this information appears to be lost in the classical limit.

In order to understand what has happened, we return to  Eq. (\ref{eq:preHSclassicalinfl}), applying the {HS}
transform to both  the $\theta_{Q}$ and $q$ variables (see Eq. (\ref{eq:multivariableGaussianidentity}) for detail). This modified influence functional
is then:
\begin{equation}
\mathcal{F}_{\rm cl}=\left\langle \exp\left(i\int_{0}^{t_{f}}{\rm d}t\ \left[-\theta_{Q}\left(t\right)\eta_{{\rm cl}}\left(t\right)+q\left(t\right)\nu_{{\rm cl}}\left(t\right)\right]\right)\right\rangle _{r},
\end{equation} 
with the $\nu_{\rm cl}$ noise defined by its correlations
\begin{align}
\left\langle {\rm \eta}_{{\rm cl}}\left(t\right){\rm \nu}_{{\rm cl}}\left(t^{\prime}\right)\right\rangle _{r}&=-2i\Theta\left(t-t^{\prime}\right)\frac{{\rm d}\gamma\left(t-t^{\prime}\right)}{{\rm d}t}, \\
\left\langle {\rm \nu}_{{\rm cl}}\left(t\right){\rm \nu}_{{\rm cl}}\left(t^{\prime}\right)\right\rangle _{r}&=0.
\end{align}
 The classical propagator for a single realisation is now expressible
as:
\begin{align}
&\tilde{U}_{{\rm cl}}=\int_{q_{0},\dot{q}_{0}}^{q_{f},\dot{q}_{f}}\mathcal{D}\bar{q}(t)\ \delta\left[m\ddot{q}\left(t\right)+V^{\prime}\left(q,t\right)-\eta_{{\rm cl}}\left(t\right)\right],\label{eq:propagatorstochasticweighting} \\
   &\mathcal{D}\bar{q}(t)= \mathcal{D}q(t){\rm e}^{i\int_{0}^{t_{f}}{\rm d}t\ q\left(t\right)\nu_{{\rm cl}}\left(t\right)}.
\end{align}
Just like in the heuristic classical limit of the SLE, the equation
of motion for an individual trajectory is a frictionless Langevin
equation. The friction component has not vanished, but its influence
on the expectations is to introduce a stochastic weighting on each
trajectory. Clearly, the equations of motion for individual trajectories
are affected by the presence or absence of a friction kernel, but
the \emph{expectations }of the two systems must be identical, provided
the appropriate stochastic weighting is used in the averaging of the
frictionless propagator. The heuristic classical limit therefore reproduces the dynamics of a frictionless Langevin system, but obscures
the resultant non-trivial weighting on trajectories for expectations required to obtain the correct averaging. 

This interpretation is not entirely satisfying, as it implies a critical
loss of information when taking the classical limit of the SLE that
must be restored with a \emph{post hoc} prescription for the weighting of
trajectories. Clearly, it would be more desirable to formulate the
SLE in such a way that its classical equation of motion corresponds
to Eq. (\ref{eq:Langevinwithfriction}) rather than Eq. (\ref{eq:classicallimitfirst}). We now detail precisely how to achieve this
reformulation. 

\subsection{Alternative SLE Classical Limit}
In order to derive a classical limit consistent with a frictional
Langevin equation, we must alter the form of the influence phase [see
Eq. (\ref{eq:quantuminfluencephase})] used to derive the SLE \emph{before} employing
the HS
transform. Returning to Eq. \ref{eq:qinfluencephase},
rather than utilising the {HS}
transformation for both $\epsilon$ and $y$, we perform it only over
$\epsilon$, such that the influence phase for a single trajectory now reads:
\begin{align}
\Phi\left[q,q^{\prime}\right] =&\frac{1}{2}\int_{0}^{t_{f}}\mathrm{d}t\int_{0}^{t_{f}}\mathrm{d}t^{\prime}\,\eta\left(t\right)\epsilon\left(t\right) +2i\int_{0}^{t_{f}}\mathrm{d}t\int_{0}^{t}\mathrm{d}t^{\prime}\,K^{I}\left(t-t^{\prime}\right)\epsilon\left(t\right)y\left(t^{\prime}\right),
\end{align}
where $\eta$ has the autocorrelation $\left\langle \eta\left(t\right)\eta\left(t^{\prime}\right)\right\rangle _{r}=K^{R}\left(t-t^{\prime}\right)$.
For the $K^{I}$ term, we integrate by parts with respect to $t^{\prime}$ to  obtain
\begin{align}
\int_{0}^{t}\mathrm{d}t^{\prime}\ K^{I}\left(t-t^{\prime}\right)y\left(t^{\prime}\right)=&\left[\gamma\left(t-t^{\prime}\right)y\left(t^{\prime}\right)\right]_{0}^{t} -\int_{0}^{t}\mathrm{d}t^{\prime}\ \gamma\left(t-t^{\prime}\right)\dot{y}\left(t^{\prime}\right).
\end{align}
The term $2i\int_{0}^{t_{f}}{\rm d}t\ \epsilon\left(t\right)\left[\gamma\left(t-t^{\prime}\right)y\left(t^{\prime}\right)\right]_{0}^{t}$ 
when expressed in the original coordinates is decoupled between the $q$ and $q^\prime$ coordinates, and just as in the classical case may be absorbed into the open system
potentials for the forward and backward propagators separately. As a result, the reduced density matrix for the
system is evolved in the following manner:
\begin{equation}
\tilde{\rho}_{t_{f}}(q;q^{\prime})=\int\mathrm{d}\bar{q}\mathrm{d}\bar{q}^{\prime}\ \tilde{U}_{{\rm eff}}\left(q,q^{\prime},t_{f};\bar{q},\bar{q}^{\prime},0\right)\tilde{\rho}_{0}\left(\bar{q};\bar{q}^{\prime}\right),
\end{equation}
with an effective propagator, $\tilde{U}_{{\rm eff}}$:
\begin{equation}
\tilde{U}_{{\rm eff}}\left(q,q^{\prime},t_{f};\bar{q},\bar{q}^{\prime},0\right)=\int\mathcal{\mathcal{D}}q\left(t\right)\mathcal{\mathcal{D}}q^{\prime}\left(t\right)\exp\left[\frac{i}{\hbar}S_{{\rm eff}}\right],
\end{equation}
defined by the effective action
\begin{align}
S_{{\rm eff}}=&\int_{0}^{t_{f}}\mathrm{d}t\,\bigg(L_{Q}\left(q\left(t\right)\right)-L_{Q}\left(q^{\prime}\left(t\right)\right)+\eta\left(t\right)\epsilon\left(t\right) -2\epsilon\left(t\right)\int_{0}^{t}\mathrm{d}t^{\prime}\ \gamma\left(t-t^{\prime}\right)\dot{y}\left(t^{\prime}\right)\bigg).
\end{align}

In this formulation, the propagator is no longer decoupled between
the forward and backward trajectories, preventing the straightforward identification of a classical limit
as in Eq. (\ref{eq:classprop}). To address this, we express $L_{Q}\left(q\left(t\right)\right)-L_{Q}\left(q^{\prime}\left(t\right)\right)$
in the sum-difference coordinates
\begin{equation}
L_{Q}\left(q\left(t\right)\right)-L_{Q}\left(q^{\prime}\left(t\right)\right)=m\dot{\epsilon}\left(t\right)\dot{y}\left(t\right)  -V\left(y\left(t\right)+\frac{\epsilon\left(t\right)}{2}\right)+V\left(y\left(t\right)-\frac{\epsilon\left(t\right)}{2}\right).
\end{equation}
To obtain the classical result, we note that the average size of the
fluctuating coordinate $\epsilon\left(t\right)$ will be proportional
to $\hbar$ \cite{KLEINERT1995}. The crucial step in obtaining the classical limit is therefore
approximating $\hbar$ as small before taking the limit:
\begin{equation}
V\left(y\left(t\right)+\frac{\epsilon\left(t\right)}{2}\right)-V\left(y\left(t\right)-\frac{\epsilon\left(t\right)}{2}\right)\approx \epsilon\left(t\right)V^{\prime}\left(y\left(t\right)\right),
\end{equation}
and $\eta\approx\eta_{{\rm cl}}$. This becomes exact in the $\hbar\to0$ limit. Note that this approach implicitly adopts the definition of the classical limit as that in which observable operators commute \cite{Cabrera2019}. Integration by parts of the
kinetic term in the effective action then yields:
\begin{align}
S_{{\rm eff}}=&\int_{0}^{t_{f}}\mathrm{d}t\ \epsilon\left(t\right)\bigg[-m\ddot{y}\left(t\right)-V^{\prime}\left(y\left(t\right)\right)\nonumber+\eta_{{\rm cl}}\left(t\right)-2\int_{0}^{t}\mathrm{d}t^{\prime}\ \gamma\left(t-t^{\prime}\right)\dot{y}\left(t^{\prime}\right)\bigg].
\end{align}
To perform the $\hbar\to0$ limit, we must examine the path integral
measure\footnote{The Jacobian for the transformation of path variables to $\epsilon$
and $y$ is unity.} in its discrete form: 
\begin{equation}
\mathcal{\mathcal{D}}y\left(t\right)\mathcal{\mathcal{D}}\epsilon\left(t\right)=\lim_{N\to\infty}\left(\frac{m}{2\pi\hbar\Delta}\right)^{N}\prod_{n}^{N}{\rm d}y_{n}{\rm d}\epsilon_{n}.
\end{equation}
Making the substitution $\theta\left(t\right)=\epsilon\left(t\right)/\hbar$,
the measure now reads
\begin{equation}
\mathcal{\mathcal{D}}y\left(t\right)\mathcal{\mathcal{D}}\theta\left(t\right)=\lim_{N\to\infty}\left(\frac{m}{2\pi\Delta}\right)^{N}\prod_{n}^{N}{\rm d}y_{n}{\rm d}\theta_{n}.
\end{equation}
Comparison to Eq. (\ref{eq:kvnmeasure}) reveals this is the KvN measure.
Furthermore, the effective propagator is now
\begin{align}
\tilde{U}_{{\rm {\rm eff}}}=&\int_{y_{0},\dot{y}_{0}}^{y_{f},\dot{y}_{f}}\mathcal{D}y(t)\mathcal{D}\theta\left(t\right)\ {\rm e}^{i\int_{0}^{t_{f}}{\rm d}t\ \theta\left(t\right)R\left(t\right)} =\int_{y_{0},\dot{y}_{0}}^{y_{f},\dot{y}_{f}}\mathcal{D}y(t)\ \delta\left[R\left(t\right)\right], \\
R\left(t\right)=&m\ddot{y}\left(t\right)+V^{\prime}\left(y,t\right)-\eta_{{\rm cl}}\left(t\right)+2\int_{0}^{t}\textrm{d}t^{\prime}\,\dot{y}(t^{\prime})\gamma\left(t-t^{\prime}\right).
\end{align}
There is now \emph{no }$\hbar$ dependence in this path integral\footnote{The effect of taking the $\hbar\to0$ limit is $y$$\left(t\right)\to q\left(t\right)$,
while the initial density matrix becomes the probability distribution
$\tilde{\rho}_{0}\left(y;\theta\right)\to\tilde{\rho}_{0}\left(q\right)$ }, and we have recovered the KvN propagator found in Eq. (\ref{eq:Langevinpropagator}).
This demonstrates that when the friction kernel is explicitly included
in the quantum mechanical path integral, the classical limit corresponds
exactly to the KvN path integral, providing a valuable consistency
check for both of these results. Furthermore, this result emphasises that in order to take a consistent classical limit of results derived with the quantum influence functional, the theory of the CIF is required to make sense of the path integral this limit produces.

	\section{Outlook\label{conclusions}}
	In this review, we have outlined the influence functional in both quantum and classical dynamics, and its use in deriving effective equations of motion for open system. While the uses of the Langevin equation hardly need enumerating, it is worth commenting on the many uses that have been found for the SLE. Its great strength is that it provides a formally exact method of evolving an open system. This has been applied in a number of areas, including quantum control \cite{Tuorila2019,PhysRevLett.107.130404}, quantum thermodynamics \cite{Kosloff2019,Stockburger2017}, EPR entanglement generation \cite{Schmidt2013} and the spin-Boson model \cite{Stockburger2002,ourpaper2,2002.07700}. Furthermore, from the perspective of numerical implementation, significant strides have been made in improving the equation's ability to scale with time, using both sampling \cite{Stockburger_2016, Schmitz2019} strategies, and blip dynamics \cite{Makri2017,Wiedmann2016}.

Beyond practical calculations, the SLE embodies some more general principles.
The inevitable breaking of time symmetry for a reduced system has been discussed
previously, but the model also touches on the necessity of memory in dynamics \cite{Koch2008}. Unlike in the classical case, there is no choice of $I(\omega)$ which will force the correlation Eq.(\ref{eq:Corr_F_Eta_Etasimple2}) to be a delta-function. Instead, it may only become truly Markovian in the limit of infinite temperature, which is itself another form of classical limit. This important feature is missed in descriptions with \emph{a priori} stochastic terms, and reflects the fact that there is no thermalisation without correlations \cite{Zhdanov2017}!

In the classical case, we have used the Koopman-von Neumann representation of classical dynamics to derive the CIF as an analogue to the Feynman-Vernon influence functional. This yields the same benefits as in the quantum case, allowing one to make direct contact between a microscopic model, and an equivalent stochastic description. This derivation may potentially be generalised in a number of ways. For example, a recent development is the incorporation of a driven environment within the CL model \cite{PhysRevE.98.012122}.
Specifically, it is possible to take a Rubin model (consisting of
two chains of oscillators coupled to a central system) \cite{PhysRev.131.964}
with a universal driving term and map this to the CL model. Using CIFs, novel stochastic representations of such a system could be derived. 

The CIF also allows one to make contact with the classical limit of quantum dynamics derived via the quantum influence functional. In the case of the SLE, the correct classical limit is found to be of a form equivalent to that derived from the CIF. 
The CIF has a more easily evaluable form than its quantum equivalent, and for this reason it may be possible to find analytic expressions for a larger class of environment models than in the quantum case. This would undoubtedly be a useful tool in the study of open-systems with anharmonic environments \cite{Bhadra2016,Bhadra_2018}. Recent progress in evaluating path integrals of singular potentials \cite{Kleinertbook} may enable the assessment of CIFs for environments with $r^{-1}$ or $r^{-2}$ potentials. Another potential avenue of extension is in the study of quantum-classical hybrids \cite{Sudarshan1976,Viennot_2018,Bondarquantumclassical}, where a quantum system interacting with a classical environment could be modeled with the use of the CIFs. 

In the quantum case, imaginary time influence functionals  have been used to describe a reduced system equilibrium state, even when the environment coupling is arbitrarily strong \cite{Moix2012}. This is important, as the stationary distribution of dissipative systems
with finite couplings has been shown to deviate from that expected under partitioned conditions \cite{PhysRevE.84.031110}, with the Gibbs distribution now being described by a "Hamiltonian of mean force" $\hat{H}_{MF}$ \cite{PhysRevLett.116.020601}. A similar result could be achieved in the classical case by using the CIF to derive an effective Hamiltonian for the reduced system. This effective Hamiltonian would then be identical to the Hamiltonian of mean force required to describe the thermal state of the reduced system. 

Finally, while influence functionals are a specific tool, the underlying formalism it is built on a statistical
interpretation of \emph{all} physics. Randomness is not an \emph{ad hoc} model addition, but an essential,
irreducible component in our description of reality. Its existence
always reflects imperfect information, whether that is due to unobserved
interactions with other systems, or a fundamentally non-commutative
algebraic structure. The surprise is that this does not just apply in the quantum realm, but the classical too. We therefore close with Max Born's articulation of this idea -- ``Ordinary mechanics must also be statistically
formulated: the determinism of classical physics turns out to be an
illusion, it is an idol, not an ideal in scientific research'' \cite{BornNobel}. 

\begin{acknowledgement}
The authors would like to thank the anonymous referees for their constructive comments, in particular for drawing our attention to the Onsager-Machlup functional, and for criticism which greatly sharpened the discussion of the interpretation of stochastic trajectories.
The authors are supported by Air Force Office of Scientific Research (AFOSR) (grant FA9550-16-1-0254; program manager Dr. Fariba Fahroo), the Army Research Office (ARO) (grant W911NF-19-1-0377; program manager Dr.~James Joseph), and Defense Advanced Research Projects Agency (DARPA) (grant D19AP00043; program manager Dr.~Joseph Altepeter). The views and conclusions contained in this document are those of the authors and should not be interpreted as representing the official policies, either expressed or implied, of AFOSR, ARO, DARPA, or the U.S. Government. The U.S. Government is authorized to reproduce and distribute reprints for Government purposes notwithstanding any copyright notation herein.

\end{acknowledgement}

\subsection*{Author Contribution Statement}

G.M. performed the derivations which were verified by D.I.B. G.M. wrote the initial manuscript text, with additional sections suggested by D.I.B. Both authors reviewed and co-edited the manuscript after the initial draft.

	\appendix

	\section{Path integrals in KvN \label{pathintegral}}
	Deriving the KvN path integral follows the
	same procedure as its quantum equivalent. Before embarking on this, it is worth considering how a change of basis is achieved in KvN. 
	
	\subsection{Basis overlaps \label{sec:bases}}
	
	In quantum mechanics, the position and momentum bases form a complementary
	pair, and it is often useful to transform between them. To do so,
	one must derive the \emph{overlap }between them. This is particularly helpful when specifying a representation of an operator in its conjugate
	basis. In the
	classical case there are four ``canonical'' sets of simultaneous
	eigenbases, these are
	\begin{align}
	\left|x,p\right\rangle, \quad 
	\left|x,\theta\right\rangle, \quad
	\left|\lambda,p\right\rangle, \quad \left|\lambda,\theta\right\rangle. \nonumber
	\end{align}

	Here we outline the procedure for deriving the overlap between two
	bases of non-commuting operators. Take two Hermitian operators $\hat{x}$
	and $\hat{y}$ with the commutation relation:
	\begin{align}
	\left[\hat{x},\hat{y}\right]=1,
	\end{align}
	it follows that
	\begin{align}
	\left[\hat{x},{\rm e}^{a\hat{y}}\right]=a{\rm e}^{a\hat{y}}.
	\end{align}
	Applying this commutator to an eigenstate of $\hat{x}$ we obtain
	\begin{align}
	\hat{x}{\rm e}^{a\hat{y}}\left|x\right\rangle =\left(x+a\right){\rm e}^{a\hat{y}}\left|x\right\rangle,
	\end{align}
	indicating ${\rm e}^{a\hat{y}}\left|x\right\rangle $ is an eigenstate
	of the $\hat{x}$ operator with eigenvalue $x+a$. From this we can
	conclude that ${\rm e}^{a\hat{y}}$ is a \emph{translation} 
	\begin{align}
	{\rm e}^{a\hat{y}}\left|x\right\rangle =\left|x+a\right\rangle .
	\end{align}
	Furthermore, it is possible give an explicit form for $\hat{y}$ in
	this basis:
	\begin{align}
	\left\langle x\right|\hat{y}\left|\psi\right\rangle =&\lim_{a\to0}\frac{1}{a}\left\langle x\right|{\rm e}^{a\hat{y}}-1\left|\psi\right\rangle \nonumber \\  =&\lim_{a\to0}\frac{1}{a}\left(\psi\left(x+a\right)-\psi\left(x\right)\right)=\frac{\partial}{\partial x}\psi\left(x\right),
	\end{align}
	i.e., $\hat{y}$ is given by $\frac{\partial}{\partial x}$ in the
	$x$ representation. 
	
	In KvN mechanics, the commutator between operators is always $i$.
	Making the assignment $i\hat{y}=\hat{\lambda}$, we can calculate
	the overlap between $\hat{x}$ and $\hat{\lambda}$: 
	\begin{align}
	\lambda\left\langle \lambda\left|x\right.\right\rangle =&\left\langle \lambda\left|i\hat{y}\right|x\right\rangle =i\frac{\partial}{\partial x}\left\langle \lambda\left|x\right.\right\rangle \\
	\implies&\left\langle \lambda\left|x\right.\right\rangle =N\left(\lambda\right){\rm e}^{-i\lambda x}.
	\end{align}
	The normalisation of the overlap is easily checked using:
	\begin{align}
	\delta\left(x-x^{\prime}\right)=&\int{\rm d}\lambda\ \left\langle x^{\prime}\left|\lambda\right.\right\rangle \left\langle \lambda\left|x\right.\right\rangle \nonumber \\ =&2\pi\left|N\left(\lambda\right)\right|^{2}\delta\left(x-x^{\prime}\right) \\
	\implies N\left(\lambda\right)=&\frac{1}{\sqrt{2\pi}}
	\end{align}
	This generically specifies the form of the overlap between eigenstates.
	Any eigenbasis of an operator is also an eigenbasis of operators it
	commutes with. Equipped with this, one may straightforwardly generate
	the following overlaps for the simultaneous eigenstates
	\begin{align}
	\left\langle x,\theta\left|x^{\prime},p\right.\right\rangle =&\frac{1}{\sqrt{2\pi}}\delta\left(x-x^{\prime}\right){\rm e}^{-i\theta p}, \label{eq:thetapoverlap} \\
	\left\langle \lambda,p\left|x,p^{\prime}\right.\right\rangle =&\frac{1}{\sqrt{2\pi}}\delta\left(p-p^{\prime}\right){\rm e}^{-i\lambda x}, \label{eq:lambdaxoverlap} \\
	\left\langle \lambda,p\left|x,\theta\right.\right\rangle =&\frac{1}{2\pi}{\rm e}^{-i\lambda x}{\rm e}^{i\theta p}.\label{eq:overlaplambdatheta}
	\end{align}
	The mathematics of specifying overlaps is generic between quantum
	and KvN mechanics, with the only generalisation arising from KvN's
	simultaneous eigenbases allowing a greater degree of freedom in representation. Equipped with this information, it is possible to represent the KvN propagator as a path integral.

	\subsection{The Propagator As A Path Integral}
		
	Take the KvN propagator,
	\begin{align}
	\hat{U}_{{\rm cl}}={\rm e}^{-it\hat{K}},
	\end{align}
	where $\hat{K}$ is given by Eq. (\ref{eq:Koopmanoperator}). In
	the phase space representation this propagator is
	\begin{align}
	U_{{\rm cl}}(x_{f},p_{f},t_{f};x_{i},p_{i},0)=\left\langle x_{f},p_{f}\left|{\rm e}^{-it_{f}\hat{K}}\right|x_{i},p_{i}\right\rangle .
	\end{align}
 Performing a Trotter splitting, this propagator may be decomposed into a product of infinitesimal propagations

		\begin{align}
		U_{{\rm cl}}(x_{f},p_{f},t_{f};x_{i},p_{i},0)=\lim_{N\to\infty}\int\textrm{d}x_{1}{\rm d}p_{1}...\textrm{d}x_{N-1}\textrm{d}p_{N-1}\prod_{j=0}^{N-1}\langle x_{j+1},p_{j+1}|{\rm e}^{-i\Delta\hat{K}}|x_{j},p_{j}\rangle.
		\end{align}
		Considering a single term in this product, we have
		\begin{align}
		\langle x_{j+1},p_{j+1}|{\rm e}^{-i\Delta\hat{K}}|x_{j},p_{j}\rangle=\langle x_{j+1},p_{j+1}|\exp\left(-i\frac{\Delta}{m}\hat{\lambda}\hat{p}\right)\exp\left(-i\Delta\hat{\theta}V^{\prime}\left(\hat{x}\right)\right)|x_{j},p_{j}\rangle,
		\end{align}
		which can be evaluated by inserting resolutions of unity
		\begin{align}
		\exp\left(-i\Delta\hat{\theta}V^{\prime}\left(\hat{x}\right)\right)|x_{j},p_{j}\rangle=&\int{\rm d}x{\rm d}\theta\ \left|x,\theta\right\rangle \left\langle x,\theta\left|x_{j},p_{j}\right.\right\rangle \exp\left(-i\Delta\theta V^{\prime}\left(x\right)\right), \label{eq:Kvnfirstterm} \\
		\langle x_{j+1},p_{j+1}|\exp\left(-i\frac{\Delta}{m}\hat{\lambda}\hat{p}\right)=&\int{\rm d}p{\rm d}\lambda\ \left\langle x_{j+1},p_{j+1}\left|\lambda,p\right.\right\rangle \exp\left(-i\frac{\Delta}{m}\lambda p\right)\left\langle \lambda,p\right|.\label{eq:KvNsecondterm}
		\end{align}
		Using the overlaps specified by Eqs. (\ref{eq:thetapoverlap}) and
		(\ref{eq:lambdaxoverlap}), we obtain for Eqs. (\ref{eq:Kvnfirstterm})
		and (\ref{eq:KvNsecondterm}): 
		\begin{align}
		\exp\left(-i\Delta\hat{\theta}V^{\prime}\left(\hat{x}\right)\right)|x_{j},p_{j}\rangle=&\frac{1}{\sqrt{2\pi}}\int{\rm d}\theta\ \left|x_{j},\theta\right\rangle \exp\left(-i\Delta\theta V^{\prime}\left(x_{j}\right)-i\theta p\right), \\
		\langle x_{j+1},p_{j+1}|\exp\left(-i\frac{\Delta}{m}\hat{\lambda}\hat{p}\right)=&\frac{1}{\sqrt{2\pi}}\int{\rm d}\lambda\ \exp\left(-i\frac{\Delta}{m}\lambda p_{j+1}+i\lambda x_{j+1}\right)\left\langle \lambda,p_{j+1}\right|.
		\end{align}
		Combining these together with Eq. (\ref{eq:overlaplambdatheta}) leads
		to the following expression for a single infinitesimal propagation
		\begin{align}
		&\langle x_{j+1},p_{j+1}|{\rm e}^{-i\Delta\hat{K}}|x_{j},p_{j}\rangle \notag= \\  &\frac{1}{\left(2\pi\right)^{2}}\int{\rm d}\lambda_{j}{\rm d}\theta_{j}\left[\ \exp\left(i\lambda_{j}\frac{\Delta}{m}\left(m\frac{x_{j+1}-x_{j}}{\Delta}-p_{j}\right)\right)\right. \left.\exp\left(i\Delta\theta_{j}\left(\frac{p_{j+1}-p_{j}}{\Delta}+V^{\prime}\left(x_{j}\right)\right)\right)\right].
		\end{align}
		Note we have added a $j$ subscript to the $\theta$ and $\lambda$
		variables in anticipation of inserting the appropriate resolutions
		of the identity. The overall propagator is therefore described by
		\begin{align}
		&U_{{\rm cl}}(x_{f},p_{f},t_{f};x_{i},p_{i},0)= \lim_{N\to\infty}\int  \prod_{j=1}^{N-1}\left(\frac{{\rm d}x_{j}}{\sqrt{2\pi}}\frac{{\rm d}p_{j}}{\sqrt{2\pi}}\frac{{\rm d}\lambda_{j}}{\sqrt{2\pi}}\frac{{\rm d}\theta_{j}}{\sqrt{2\pi}}\right)  \nonumber \\  &\times  \exp\left(i\Delta \sum_{j=0}^{N-1}\left[\lambda_{j}\left(\frac{x_{j+1}-x_{j}}{\Delta}-\frac{p_{j}}{m}\right)+\theta_{j}\left(\frac{p_{j+1}-p_{j}}{\Delta}+V^{\prime}\left(x_{j}\right)\right)\right]\right).
		\end{align}
		
		In the limit we can once again describe this with a functional notation (although we have cheated and moved directly to describing a time-dependent
			potential, which can be justified in the same way as in the quantum
			case \cite{TechniquesApplicationsPathIntegration})
		\begin{align}
		&U_{{\rm cl}}\left(x_{f},p_{f},t_{f};x_{i},p_{i},0\right)=\int_{x_{i},p_{i}}^{x_{f},p_{f}}\mathcal{D}x\mathcal{D}p\mathcal{D}\lambda\mathcal{D}\theta\ {\rm e}^{iR}, \\
		&R=\int_{0}^{t_{f}}{\rm d}t\ \left[\lambda\left(t\right)\left(\dot{x}\left(t\right)-\frac{p\left(t\right)}{m}\right)+\theta\left(t\right)\left(\dot{p}\left(t\right)+V^{\prime}\left(x(t),t\right)\right)\right].\label{eq:Kpathintexponent}
		\end{align}
		The functional measure for each path variable is 
		\begin{align}
		\mathcal{D}f=\lim_{N\to\infty}\prod_{n}^{N}\frac{{\rm d}f_{n}}{\sqrt{2\pi}}
		\end{align}
		and compared to the quantum path integral, there is no factor of $i$
		causing the measure to fluctuate. For this reason, the KvN path integral
		is well behaved in the continuous limit.
		
		The raw form of the KvN propagator is not particularly illuminating,
		but the integration over the non-observable variables $\lambda$ and $\theta$ represents a product
		of delta functionals enforcing Hamilton's equations. We can see this
		most easily by returning to the discrete formulation. Specifically,
		consider the integration over $\lambda_{j}$
		\begin{align}
		\int{\rm d}\lambda_{j}\ \exp\left(i\lambda_{j}\frac{\Delta}{m}\left(m\frac{x_{j+1}-x_{j}}{\Delta}-p_{j}\right)\right) =\frac{m}{\Delta}2\pi\delta\left(m\frac{x_{j+1}-x_{j}}{\Delta}-p_{j}\right).
		\end{align}
		If this delta function is now integrated with respect to $p_{j}$,
		the propagator may be expressed with a reduced number of path variables.
		The functional measure is now
		\begin{align}
		\mathcal{D}x\mathcal{D}\theta=\lim_{N\to\infty}\left(\frac{m}{2\pi\Delta}\right)^{N}\prod_{n}^{N}{\rm d}x_{n}{\rm d}\theta_{n}, \label{eq:kvnmeasureappendix}
		\end{align}
		while the propagator itself is
		\begin{align}
		U_{{\rm cl}}\left(x_{f},p_{f},t_{f};x_{i},p_{i},0\right)=\int_{x_{i},\dot{x}_{i}}^{x_{f},\dot{x}_{f}}\mathcal{D}x\mathcal{D}\theta\ \exp\left[i\int_{0}^{t_{f}}{\rm d}t\ \theta\left(t\right)\left(m\ddot{x}\left(t\right)+V^{\prime}\left(x(t),t\right)\right)\right].\label{eq:Koopmanpropagatorappendix}
		\end{align}

	\section{The Driven Harmonic oscillator \label{sec:Oscillator}}
In order to evaluate the CIF for the CL model, we require the solution to 
\begin{equation}
m\ddot{x}(t)=-m\omega^{2}x(t)+f(t).\label{eq:forcedoscillator}
\end{equation}
Solving this equation is not entirely trivial, but can be accomplished
in a variety of ways (A Green's function approach is often used here).
In the interest of novelty we shall take a slightly different approach, by re-expressing Eq. (\ref{eq:forcedoscillator})
as a first-order matrix equation
\begin{equation}
m\dot{X}(t)=mAX+F(t),
\end{equation}
\begin{equation}
X=\begin{pmatrix}x(t)\\
\dot{x}(t)
\end{pmatrix}, \quad A=\begin{pmatrix}0 & 1\\
-\omega^{2} & 0
\end{pmatrix}, \quad F=\begin{pmatrix}0\\
f(t)
\end{pmatrix}.
\end{equation}
Solving this equation with the integrating factor $\exp(At)$ yields
\begin{equation}
X(t)=\textrm{e}^{-At}X(0)+\frac{1}{m}\int_{0}^{t}\textrm{d}s\,\textrm{e}^{-A(t-s)}f(s).\label{eq:harmeqnmotion}
\end{equation}
This solution can be recast by expanding the matrix exponentials.
This first requires the evaluation of $A^{n}$:
\begin{equation}
A^{n}=\begin{cases}
(-1)^{n/2}\omega^{n}\begin{pmatrix}1 & 0\\
0 & 1
\end{pmatrix} & n\textrm{{\,\ even}}\\
(-1)^{(n-1)/2}\omega^{n-1}\begin{pmatrix}0 & 1\\
-\omega^{2} & 0
\end{pmatrix} & n\textrm{{\,\ odd}}
\end{cases}
\end{equation}
which can be used to rearrange the exponential expansion into odd
and even terms
\begin{align}
\textrm{e}^{-At}=&\sum_{n=0}^{\infty}\left\{ \frac{(At)^{2n}}{(2n)!}+\frac{(At)^{2n+1}}{(2n+1)!}\right\} =\cos(\omega t)\begin{pmatrix}1 & 0\\
0 & 1
\end{pmatrix}+\sin(\omega t)\begin{pmatrix}0 & \omega^{-1}\\
-\omega & 0
\end{pmatrix}.
\end{align}
Substituting this into Eq. (\ref{eq:harmeqnmotion}) we obtain:
\begin{align}
\begin{pmatrix}x(t)\\
\dot{x}(t)
\end{pmatrix}&=\cos(\omega t)\begin{pmatrix}x(0)\\
\dot{x}(0)
\end{pmatrix}+\sin(\omega t)\begin{pmatrix}\dot{x}(0)/\omega\\
-\omega x(0)
\end{pmatrix} 
&+\int_{0}^{t}\textrm{d}s\,\cos(\omega(t-s))\begin{pmatrix}0\\
f(s)
\end{pmatrix} \nonumber \\ &+\frac{1}{m}\int_{0}^{t}\textrm{d}s\,\sin(\omega(t-s))\begin{pmatrix}f(s)/\omega\\
0
\end{pmatrix}.\label{eq:harmoniceqnmotion}
\end{align}
A nice feature of this method is that once the equation of motion
is obtained reading off the top row, there is a free consistency check
that its derivative is equal to the bottom row. Using $A$ and $B$
for constants and grouping terms produces
\begin{align}
x(t) =&A\sin\left(\omega t\right)+B\sin\left(\omega(t_{f}-t)\right)+\frac{1}{m\omega}\int_{0}^{t}\textrm{d}s\,f(s)\sin\left(\omega(t-s)\right), \label{eq:forcedmotion} \\
\dot{x}(t)=&A\omega\cos\left(\omega t\right)-B\omega\cos\left(\omega(t_{f}-t)\right)+\frac{1}{m}\int_{0}^{t}\textrm{d}s\,f(s)\cos\left(\omega(t-s)\right).
\end{align}

	\section{The Hubbard Stratonovich Transformation \label{HStransform}}
	
	Consider a complex Gaussian distribution $W(\vec{z})$:
		\begin{equation}
	W[\eta_{1},\eta_{1}^{*}, \ldots, \eta_{N},\eta_{N}^{*}] =C \exp\left[-\frac{1}{2}\vec{z}^{T}\Phi\vec{z}\right].
	\end{equation}
	Here $\vec{z}$ is the vector of all the complex variables and their
	conjugates, with individual elements labeled as $z_{i}^{\alpha}$
	\begin{align}
	\vec{z}=\left(\begin{array}{c}
	\vec{z}_{1}\\
	\vec{z}_{2}\\
	\vdots\\
	\vec{z}_{N}
	\end{array}\right), \quad
	 \vec{z}_{i}=\left(\begin{array}{c}
	z^1_{i}\\
	z^2_{i}
	\end{array}\right)=\left(\begin{array}{c}
	\eta_{i}\\
	\eta_{i}^{*}
	\end{array}\right).
	\end{align}
The Fourier transform of this distribution is:
	\begin{align}
	\kappa(k_{1},\ldots,k_{N}) & =\int\mathrm{d}\vec{z}\ \ W(\vec{z})\mathrm{e}^{i\vec{z}^{T}\vec{k}}= \int\mathrm{d}\vec{z}\ \exp\left[-\frac{1}{2}\vec{z}^{T}\Phi\vec{z}+i\vec{z}^{T}\vec{k}\right].\label{eq:functionalaverage}
	\end{align}
	Evaluating the Fourier transform is simply a case of completing the
	square of the exponent and produces
	\begin{equation}
	\kappa[\vec{k}]=\mathrm{e^{-\frac{1}{2}\vec{k}^{T}\Phi^{-1}\vec{k}}}.
	\end{equation}
	The exponent may be expanded in terms of the random
	variable correlations:
	\begin{equation}
	\vec{k}^{T}\Phi^{-1}\vec{k}=\sum_{ij\alpha\beta}k_{i}^{\alpha}\left\langle z_{i}^{\alpha}z_{j}^{\beta}\right\rangle_r k_{j}^{\beta}.
	\end{equation}
	
We now order each random variable by a parameter $t_{i}$, where the value of each parameter is evenly spaced by a gap $\Delta$. If $t_{N}=t_{f}$, $t_{0}=0$, the gap is given by $\Delta=\frac{t_{f}}{N}$.
	Defining now a single process $z_{i}^{\alpha}=z^{\alpha}(t_{i})$
	$k_{i}^{\alpha}=\Delta k^{\alpha}(t_{i})$, we take the continuum
	limit $N\to\infty$. In this limit, vector and matrix products become
	integrals
	\begin{align}
	\sum_{i}z_{i}^{\alpha}k_{i}^{\alpha} =\sum_{i}\Delta z^{\alpha}(t_{i})k^{\alpha}(t_{i})\to\int_{0}^{t_{f}}\mathrm{d}t\ z^{\alpha}(t)k^{\alpha}(t), \\
	\sum_{ij}k_{i}^{\alpha}\left\langle z_{i}^{\alpha}z_{j}^{\beta}\right\rangle_r k_{j}^{\beta} =\sum_{i,j}\Delta^{2}k^{\alpha}(t_{i})A^{\alpha\beta}\left(t_{i},t_{j}\right)k^{\beta}(t_{j}) \nonumber \\  \to\int_{0}^{t_{f}}\int_{0}^{t_{f}}\mathrm{d}t\textrm{d}t^{\prime}\ \ k^{\alpha}(t)A^{\alpha\beta}\left(t,t^{\prime}\right)k^{\beta}(t^{\prime}).
	\end{align}
	Here the matrix $A^{\alpha\beta}$ is defined in relation to $\Phi^{\alpha\beta}$ as follows:
	\begin{align}
	\sum_{\beta}\int_{0}^{t_{f}}dt^{\prime}\ \Phi^{\alpha\beta}(t,t^{\prime})A^{\beta\gamma}(t^{\prime},t^{\prime\prime})=&\delta(t-t^{\prime\prime})\delta_{\alpha\gamma}, \\
	A^{\alpha\beta}\left(t,t^{\prime}\right)=&\left\langle z_{i}^{\alpha}\left(t\right)z_{j}^{\beta}\left(t^{\prime}\right)\right\rangle _{r}.
	\end{align}
	Having taken the continuous limit, the measure for the integration
	is now akin to a path integral, as $\lim_{N\to\infty}\prod_{i}^{N}\mathrm{d}z_{i}^{\alpha}\to\mathcal{D}z^{\alpha}(\tau)$.
	In the continuous limit, the Fourier transform $\kappa$ becomes
	\begin{equation}
	\kappa(\vec{k}(t_{f}))=\exp\left[-\frac{1}{2}\sum_{\alpha\beta}\int_{0}^{t_{f}}\int_{0}^{t_{f}}\mathrm{d}t\textrm{d}t^{\prime}\ k^{\alpha}(t)A^{\alpha\beta}\left(t,t^{\prime}\right)k^{\beta}(t^{\prime})\right].\label{eq:fouriertransformhubstron}
	\end{equation}
	Remembering the original definition of $\kappa$ in Eq. (\ref{eq:functionalaverage}),
	it is possible to interpret this not just as a Fourier transform but
	as a functional average
	\begin{equation}
	\kappa(\vec{k}(t_{f}))=\left\langle \exp\left[i\sum_{\alpha}\int_{0}^{t_{f}}\mathrm{d}t\ \ z^{\alpha}(t)k^{\alpha}(t)\right]\right\rangle _{r}.
	\end{equation}
	Importantly, the relationship between the $k^{\alpha}$ is \emph{not}
	constrained in the same way as the variables $z^{\alpha}$ are. This
	means we are free to choose what, if any, functional dependence there
	is between $k^{1}(t)$ and $k^{2}(t)$. 
	
	Putting all of this together gives us the \emph{Hubbard-Stratonovich
	}(HS) transformation\footnote{Invented by Stratonovich, popularised outside the USSR by Hubbard.}:
	\begin{align}
	&\left\langle \exp\left[i\sum_{\alpha}\int_{0}^{t_{f}}\mathrm{d}t\ \ z^{\alpha}(t)k^{\alpha}(t)\right]\right\rangle _{r} = \nonumber \\
	&\exp\left[-\frac{1}{2}\sum_{\alpha\beta}\int_{0}^{t_{f}}\int_{0}^{t_{f}}\mathrm{d}t\textrm{d}t^{\prime}\ \ k^{\alpha}(t)\left\langle z^{\alpha}(t)z^{\beta}(t^{\prime})\right\rangle k^{\beta}(t^{\prime})\right]. \label{eq:HSindentity}
	\end{align}
which is easily generalised to multivariate processes
	\begin{align}
	&\left\langle \exp\left[i\sum_{i\alpha}\int_{0}^{t_{f}}\mathrm{d}t\ \ z_{i}^{\alpha}(t)k_{i}^{\alpha}(t)\right]\right\rangle _{r}= \nonumber \\
	&\exp\left[-\frac{1}{2}\sum_{ij\alpha\beta}\int_{0}^{t_{f}}\mathrm{d}t\int_{0}^{t_{f}}\textrm{d}t^{\prime}\ \ k_{i}^{\alpha}(t)\left\langle z_{i}^{\alpha}(t)z_{j}^{\beta}(t^{\prime})\right\rangle k_{j}^{\beta}(t^{\prime})\right]. \label{eq:multivariableGaussianidentity}
	\end{align}

	\bibliographystyle{aipnum4-1}
	\addcontentsline{toc}{section}{\refname}\bibliography{refs}

\end{document}